\documentclass[apj]{emulateapj}

\usepackage{color}
\usepackage{natbib}
\usepackage{amsmath}
\usepackage{graphicx}
\usepackage{graphicx,epstopdf}
\usepackage{enumerate}
\usepackage[bookmarks=true, pdfstartview={FitH},hyperfootnotes=false]{hyperref}

\shortauthors{Y. Uprety, et al.}

\begin{document}	
\title{Solar Wind Charge Exchange contribution to the ROSAT All Sky Survey Maps}


\author{Y. Uprety\altaffilmark{1}, M. Chiao\altaffilmark{2}, M. R. Collier\altaffilmark{2}, T. Cravens\altaffilmark{3}, M. Galeazzi\altaffilmark{1}, D. Koutroumpa\altaffilmark{4}, K. D. Kuntz\altaffilmark{5}, R. Lallement\altaffilmark{6}, S. T. Lepri\altaffilmark{7}, W. Liu\altaffilmark{1}, D. McCammon\altaffilmark{8}, K. Morgan\altaffilmark{8}, F. S. Porter\altaffilmark{2}, K. Prasai\altaffilmark{9}, S. L. Snowden\altaffilmark{2}, N. E. Thomas\altaffilmark{2}, E. Ursino\altaffilmark{1}, B. M. Walsh\altaffilmark{10}}

%
%
%
%
%
%
%
%

\altaffiltext{1}{Department of Physics, University of Miami, Coral Gables, FL, 33124, U.S.A.}
\altaffiltext{2}{NASA Goddard Space Flight Center, Greenbelt, MD,  20771, U.S.A.}
\altaffiltext{3}{Department of Physics and Astronomy, University of Kansas, Lawrence, KS 66045, U.S.A.}
\altaffiltext{4}{LATMOS/IPSL, UVSQ Université Paris-Saclay, UPMC Université Paris 06, CNRS, Guyancourt, France}
\altaffiltext{5}{The Henry A. Rowland Department of Physics and Astronomy, Johns Hopkins University, Baltimore, MD 21218, U.S.A.}
\altaffiltext{6}{GEPI Observatoire de Paris, CNRS, Universite Paris Diderot, 92190, Meudon, France}
\altaffiltext{7}{Department of Atmospheric, Oceanic, and Space Sciences, University of Michigan, Ann Arbor, MI 48109, U.S.A.}
\altaffiltext{8}{Department of Physics, University of Wisconsin, Madison, WI 53706, U.S.A.}
\altaffiltext{9}{Department of Physics and Engineering, California State University, Bakersfield, CA 93311, U.S.A.}
\altaffiltext{10}{Department of Mechanical Engineering, Boston University, Boston, MA 02215, U.S.A.}

\begin{abstract}
{\it DXL} (Diffuse X-ray emission from the Local Galaxy) is a sounding rocket mission designed to estimate the contribution of Solar Wind Charge eXchange (SWCX) to the Diffuse X-ray Background (DXB) and to help determine the properties of the Local Hot Bubble (LHB). The detectors are large-area thin-window proportional counters with a spectral response similar to that of the PSPC used in the {\it ROSAT} All Sky Survey (RASS). A direct comparison of {\it DXL} and RASS data for the same part of the sky viewed from quite different vantage points in the Solar system and the assumption of approximate isotropy for the Solar wind allowed us to quantify the SWCX contribution to all 6 RASS bands (R1-R7, excepting R3). We find that the SWCX contribution  at $l=140^\circ, b=0^\circ$, where the {\it DXL} path crosses the Galactic plane is  $33\%\pm6\% (statistical)\pm12\%(systematic)$ for R1, $44\%\pm6\%\pm5\%$ for R2, $18\%\pm12\%\pm11\%$ for R4, $14\%\pm11\%\pm9\%$ for R5, and negligible for R6 and R7 bands.  Reliable models for the distribution of neutral H and He in the Solar system permit estimation of the contribution of interplanetary SWCX emission over the the whole sky and correction of the RASS maps. We find that the average SWCX contribution in the whole sky is $26\%\pm6\%\pm13\%$ for R1, $30\%\pm4\%\pm4\%$ for R2, $8\%\pm5\%\pm5\%$ for R4, $6\%\pm4\%\pm4\%$ for R5, and negligible for R6 and R7.
\end{abstract}



\section{Introduction}
\label{introduction}

The {\it ROSAT} All-Sky Survey (RASS) provides the best maps of the soft diffuse X-ray background \citep{Snowden1997}. These maps are consistent with previous all-sky measurements of the diffuse background but have far superior statistics and angular resolution, and after more than 20 years are still the benchmark for all diffuse emission studies at energies less than 2 keV. 

It was widely accepted that the diffuse X-ray flux observed in the $\sim$1/4~keV (R12) band originated largely in a $\sim$100 parsec (pc) scale million-degree interstellar plasma in the Local Hot Bubble (LHB) with a significant contribution at high latitudes from patches of million-degree emission in the Galactic Halo \citep{McCammon1990,KuntzSnowden2000,BellmVaillancourt2005}.  The 3/4~keV (R45) band  arises in $1.5-3\times 10^6$~K plasma located more extensively throughout the Galactic disk and halo, with 30-50\% of the flux at intermediate and high latitudes due to AGN and a modest but poorly-determined contribution due to stars close to the Galactic plane and toward the center \citep{Snowden1997, Mushotzky2000}.  Shadowing studies using molecular clouds located near the outer boundary of the LHB showed that most of the observed R45 flux originated beyond these clouds, and that the several percent that was ``local'' was concentrated in the OVII lines near the low-energy limit of the band \citep{snowden93,Smith2007,Henley2008a,Gupta2009b}.  The X-ray flux in the R67 bands above 1~keV is thought to be almost entirely from AGN.

During the course of the six-month {\it ROSAT} survey it became apparent there was a significant source of background that was strongly variable on a time scale of several hours to a couple of days in the R12 and R45 bands \citep{Snowden1994a}.  
The source was unknown, but since the time variation was short compared to the observing time for each part of the sky, these
``long-term enhancements'' (LTEs) could be empirically removed.  
The good agreement with the zero points of earlier surveys gave some confidence that there were no large residuals \citep{Snowden1995a}.  Much later, after the discovery of X-ray emission from comets \citep{lisse96}, it was recognized that the LTEs were largely due to Solar Wind Charge eXchange (SWCX) in the outermost parts of the Earth's geocorona (Geocoronal SWCX)\citep{Cox1998, cravens2001}. SWCX refers to the process by which the heavy ions of the solar-wind pick up electrons from neutrals (mostly H or He) to emit EUV or low energy X-rays \citep{cravens1997}. It was soon realized that there would also be X-rays from SWCX on interstellar neutral H and He passing through interplanetary space (Heliospheric SWCX) \citep{Cox1998}.  However, the much slower time variations in this component due to the extended spatial distribution of the target neutrals largely precluded its being removed in the survey analysis.

Despite numerous efforts, an accurate estimation of the heliospheric SWCX contribution has been hindered by the poorly known cross sections for producing the many X-ray lines from SWCX, limited data on heavy ion fluxes in the Solar wind, and the general spectral similarity of SWCX and thermal emission.   The best current calculations show that SWCX could generate anywhere from 25\%  to 100\% of the R12 band emission in directions of the minimum observed flux, e.g., the Galactic plane \citep{Cravens2000,Koutroumpa2009b}.  As a result, studies of soft X-ray emission from the ISM and beyond were plagued by the uncertainty of an unknown SWCX contribution, and even the existence of the LHB was held by some to be in doubt (e.g., \citet{WelshShelton2009}).  

The DXL mission was designed to reduce these uncertainties by direct measurement of the interplanetary SWCX contribution \citep{Galeazzi2011,Galeazzi2012a,Thomas2013,Galeazzi2014, Uprety2015}. This was done by comparing the X-ray flux measured during a scan across a high density feature in the interstellar neutral distribution called the ``helium focusing cone'' with observations of the same part of the sky made 22 years and 3 months earlier during the {\it ROSAT} survey, when the lines of sight passed through more typical parts of the solar system.  Given well-established models for the 3-dimensional distribution of neutral H and He in interplanetary space \citep{Koutroumpa2006,Lallement1985a,Lallement1985b,Lallement2004}, all of the highly uncertain atomic parameters  for production of many X-ray lines by charge exchange and the poorly known fluxes of various heavy ions  in the solar wind could be rolled into a single parameter to be fit to the data (see \S~3).  An additional parameter was introduced to allow for differences in the Solar wind flux between the {\it ROSAT} and {\it DXL} observations.  The model has to assume isotropy of the Solar wind, but both {\it ROSAT} and {\it DXL} observations were made close to maximum of the Solar cycle where this is not unreasonable, particularly for the small region near the ecliptic plane where the observations were made.  

Our initial analysis of these measurements indicates that SWCX contributes about 40\% of the R12 band on the Galactic plane near $l=140^\circ$, where the R12 flux is close to its minimum value \citep{Galeazzi2014}.  In this paper we extend these results to the other RASS bands, and use the best fit coefficients for the atomic and Solar wind characteristics to predict interplanetary SWCX contributions to the RASS maps over the entire sky.   In \S~2 we briefly describe the characteristics of the {\it DXL} mission, in \S~3 we focus on the data analysis, \S~4 contains the result of the investigation, and conclusions are discussed in \S~5.

\begin{figure}
\figurenum{1}
\plotone{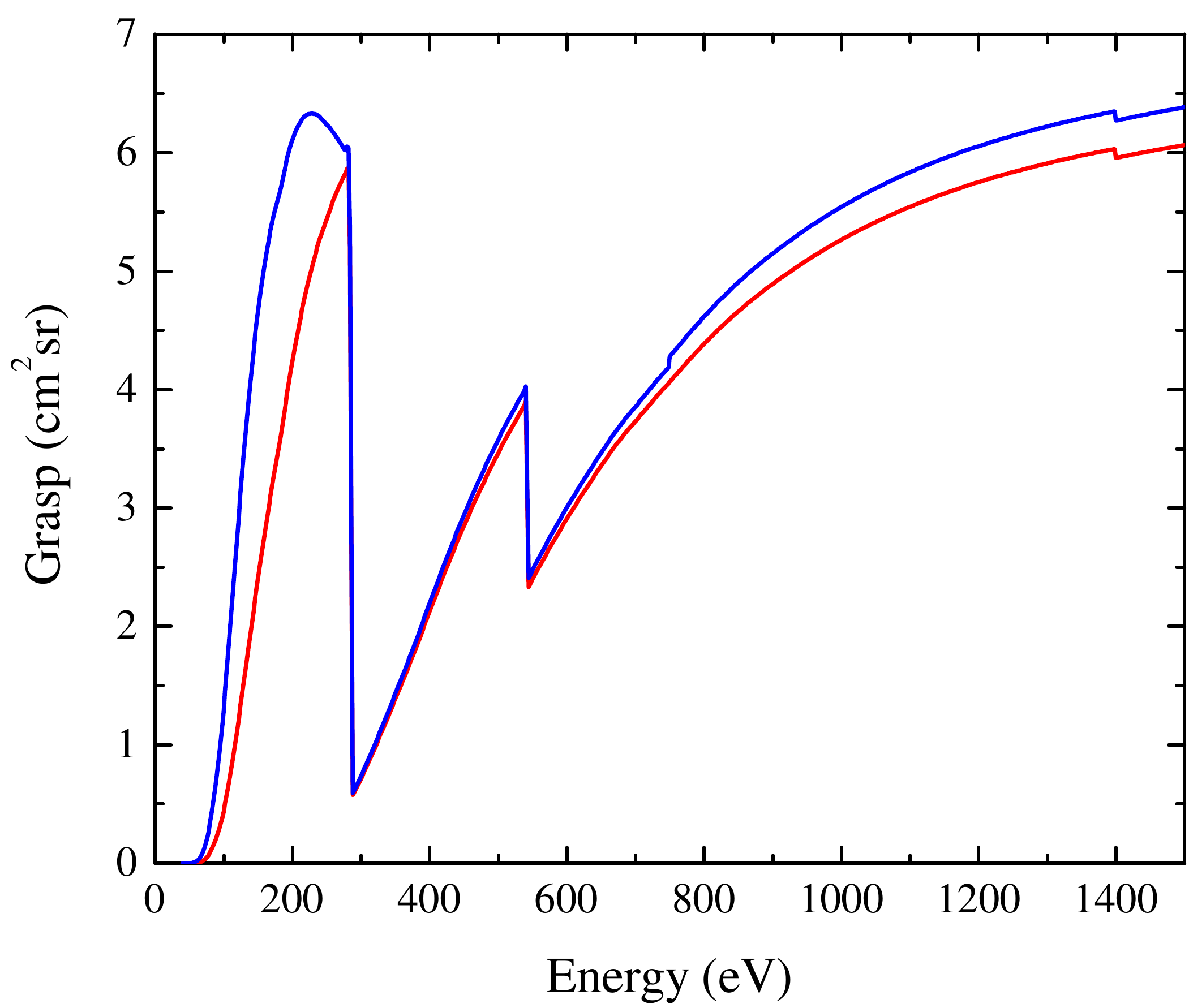}
\caption{ DXL CI (red) and CII (blue) grasp (effective area times solid angle).} 
\label{fig: graspall}
\end{figure}

\section{DXL}
\label{DXL}

The technical details of the {\it DXL} mission were extensively described in \cite{Galeazzi2011}. To summarize, {\it DXL} is a sounding rocket mission designed to measure the contributions of the LHB and SWCX through the spatial signature of their emission. The main instrument on the payload is {\it DXL}, which consists of two large area proportional counters mounted on an aluminum frame and supported by the rocket skin. The instrument was inherited and refurbished from the Wisconsin All Sky Survey program \citep{McCammon1983}. 
The payload also carried the first prototype of a wide field-of-view soft X-ray camera using micropore reflector technology \citep{Collier2015}. {\it DXL} was successfully launched from White Sands Missile Range in New Mexico on December 12, 2012.

The {\it DXL} instrument is composed of two co-aligned, large area ($1,000$~cm$^2$ physical area per counter) proportional counters (CI and CII) that provide excellent counting statistics when scanning the sky in the limited observing time of a sounding rocket flight. The grasps of the two DXL counters are shown in Figure~\ref{fig: graspall}. When coupled with the optics ($6.5^\circ$ FWHM mechanical collimators, electron blocking magnets, and carbon-based thin windows with supporting mesh for the proportional counter gas), the grasp (effective area-solid angle product) of each counter in the 1/4~keV range is $\sim$6~cm$^2$~sr.

The counters use a wire-wall design and are filled with P10 gas (90\% Argon, 10\% methane) at about 800 Torr. The X-rays are mechanically collimated using 2.5 cm thick, 3 mm cell honeycomb. Ceramic magnets embedded in the collimator provide a magnetic field of 150~Gauss within and above the collimators which intercept electrons that could either generate X-rays when impacting the collimator structure, or mimic X-ray events in the detectors. The detector employs a three-sided veto for cosmic-ray rejection using anodes at the sides and back of each main counter.

\begin{figure}
\figurenum{2}
\plotone{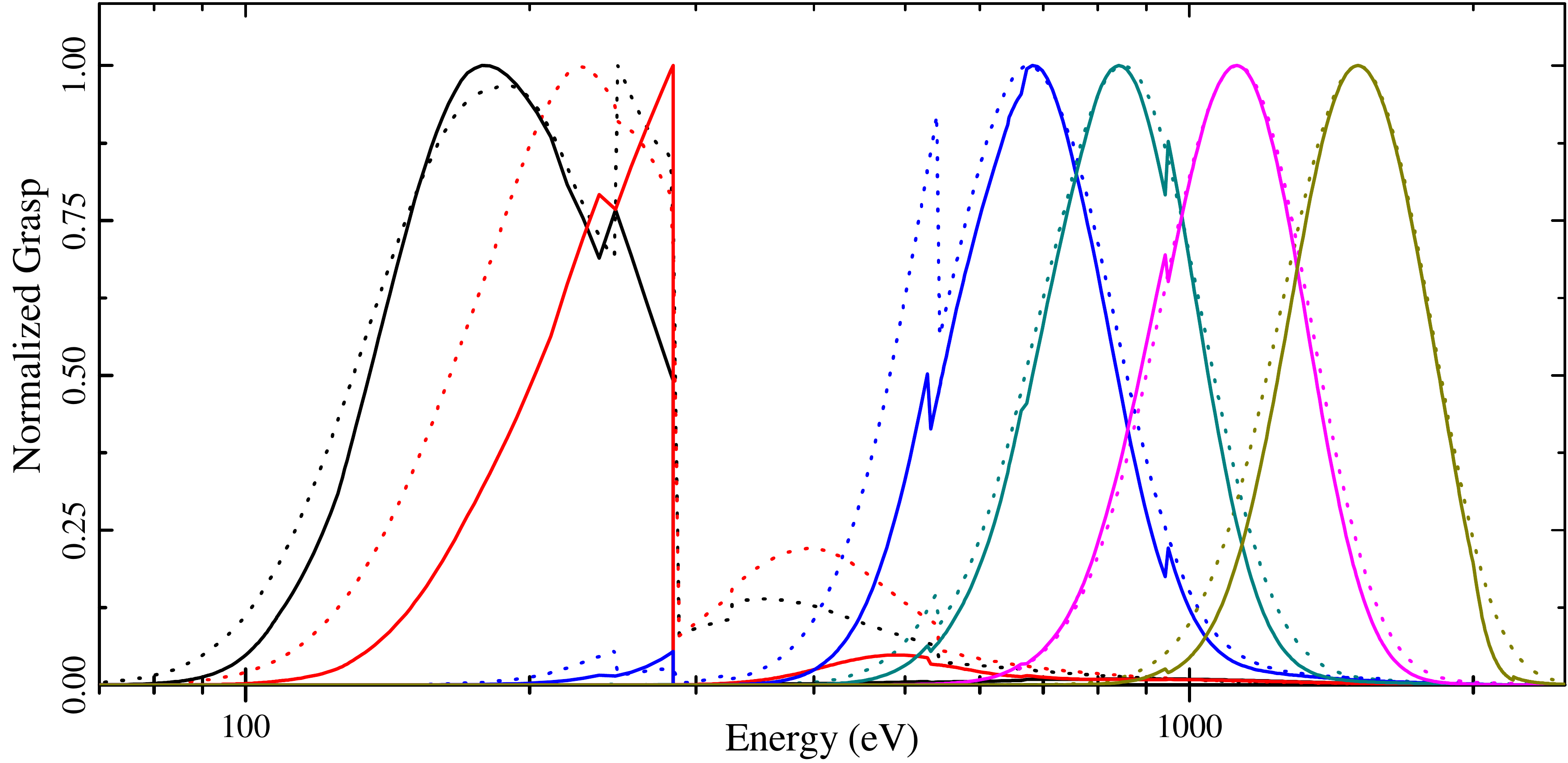}
\caption{We matched DXL bands (dotted) as closely as possible to the ROSAT bands (solid) for direct comparison. The {\it DXL} bands are labeled as {\color{black} D1 (black)}, {\color{red}D2 (red)}, {\color{green}D4 (green)}, {\color{blue}D5 (blue)}, {\color{cyan}D6 (cyan)} and {\color{magenta}D7 (magenta)} in correspondence to {\it ROSAT}'s {\color{black} R1 (black)}, {\color{red}R2 (red)}, {\color{green}R4 (green)}, {\color{blue}R5 (blue)}, {\color{cyan}R6 (cyan)} and {\color{magenta}R7 (magenta)} bands. The bands are defined by the grasp (a product of effective area to solid angle in cm$^2$~sr as a function of energy.} 
\label{fig: graspdxlrosat}
\end{figure} 

One side of each counter, facing the mechanical collimators, is covered with a thin X-ray window (25~cm~$\times$~50~cm). For the first {\it DXL} flight, we used thin (80-90~mg~cm$^{-2}$) plastic windows supported by a 25~$\mu$m thick, 100 lines-per-inch nickel mesh to retain the counter gas while allowing soft X-ray photons to enter the counter. The windows are composed of Formvar (the registered trade name of the polyvinyl formal resin produced by Monsanto Chemical Company) with an additive, Cyasorb UV-24 made by American Cyanamide, to absorb stellar ultraviolet photons that could otherwise generate a large non-X-ray background. The windows were originally manufactured for the DXS shuttle mission \citep{Sanders2001} and DXL used the spares that remained preserved. The windows were completely re-calibrated for the DXL mission. While the energy resolution of the proportional counters is relatively poor, the carbon edge at 0.284 keV in the filter response allows a clear separation of the events into different bands, roughly corresponding to those used for the RASS. Figure~\ref{fig: graspdxlrosat} shows the normalized {\it DXL} (D1-D7) and RASS (R1-R7) bands for comparison.

\begin{figure}
\figurenum{3}
\plotone{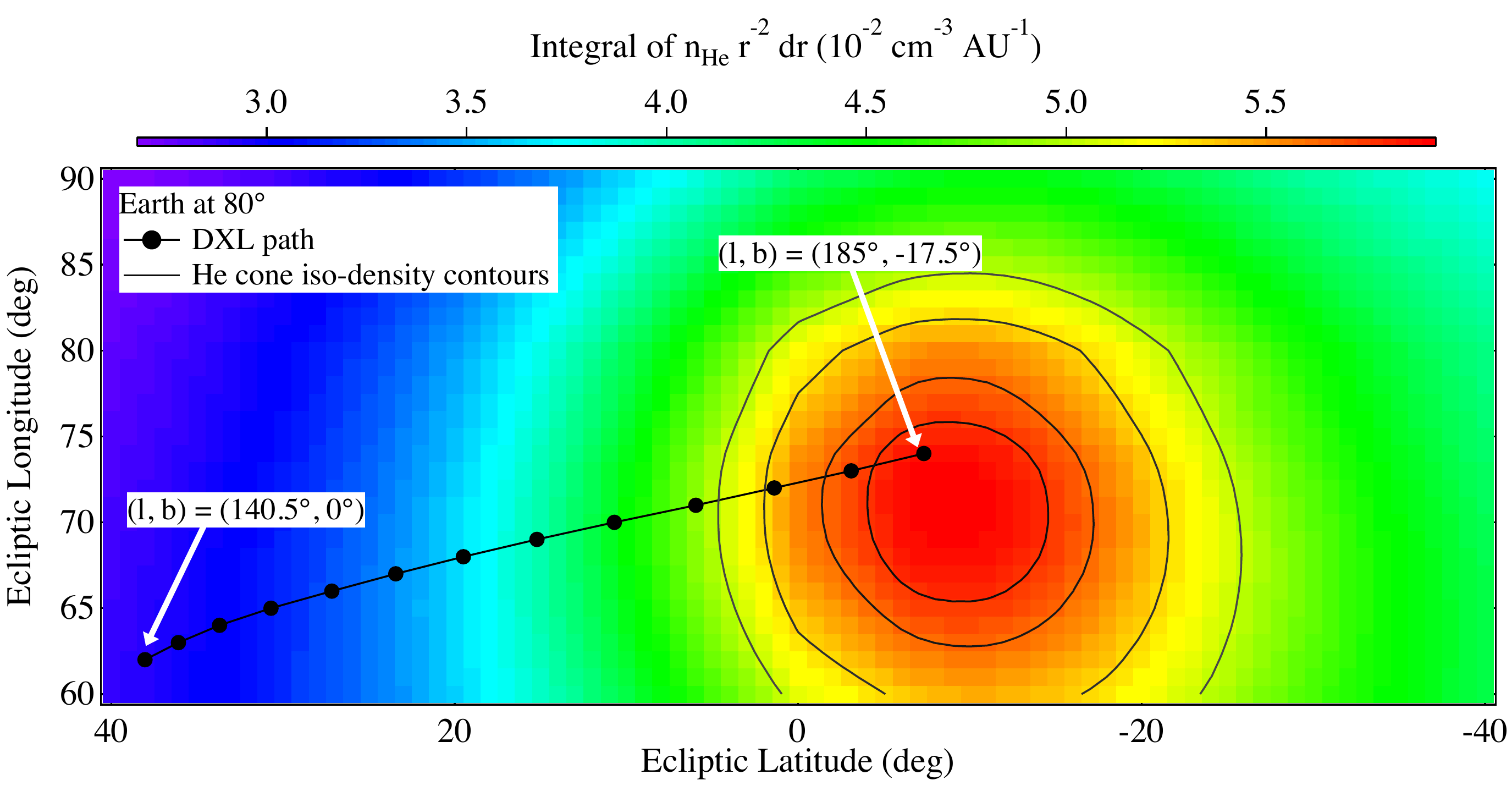}
\caption{{\it DXL} scan path into the He Focusing Cone. The color map shows the central region of the cone, with the colorscale corresponding to the He density distribution weighted by one over the distance from the Sun squared ($\frac{1}{R^2}$) to reflect the dilution of the solar wind as it flows outwards.}
\label{fig: scanpath}
\end{figure}

\begin{figure}
\figurenum{4}
\plotone{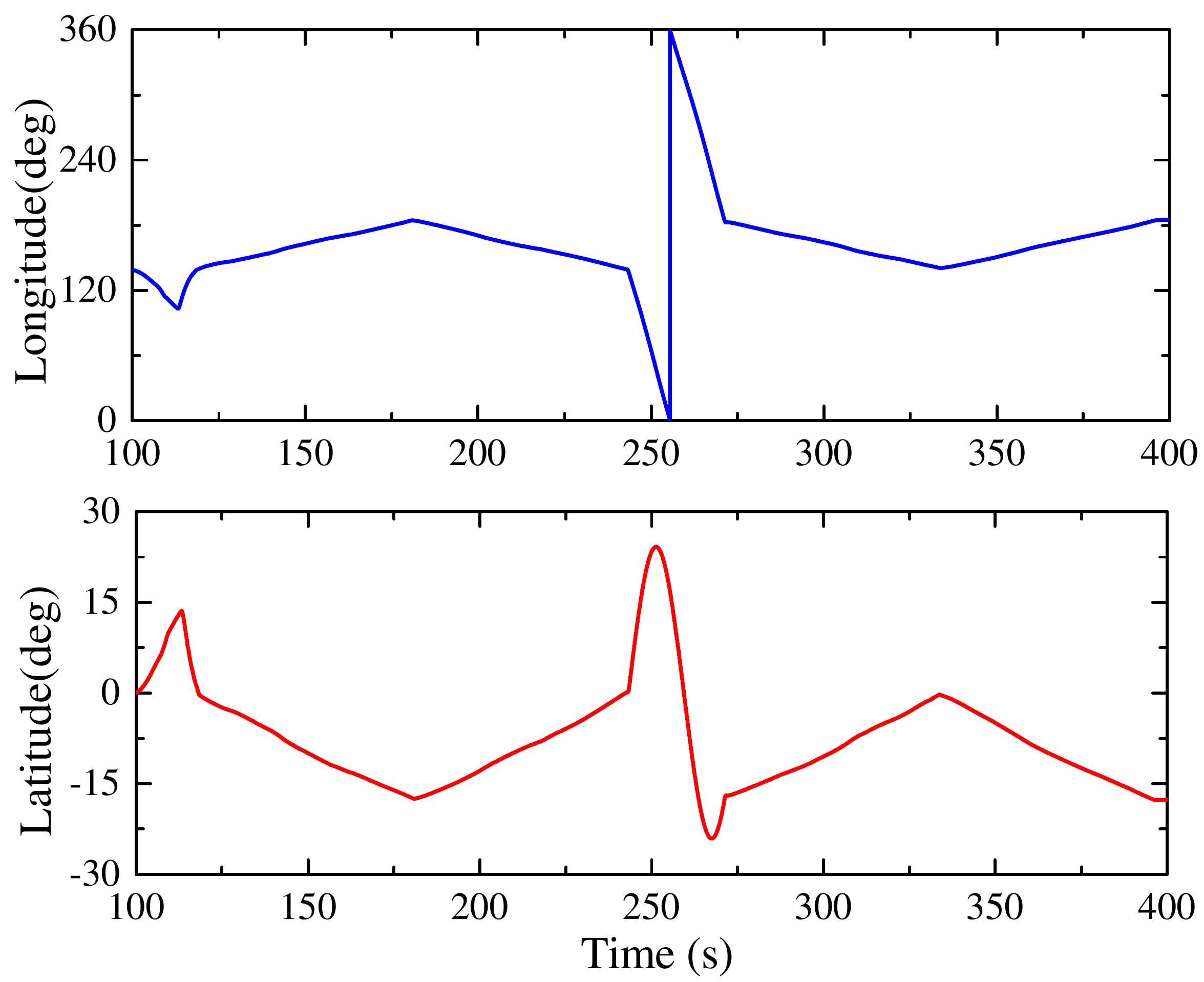}
\caption{DXL scan path shown as a function of time in Galactic co-ordinates.}
\label{fig: acs}
\end{figure}

To separate the SWCX contribution from the other components of the DXB, the {\it DXL} mission uses the spatial signature of SWCX emission due to the ``He focusing cone'', a higher neutral He density region downwind of the Sun \citep{Michels2002}.  
Neutral interstellar gas flows at $\sim$25~km~s$^{-1}$ through the solar system due to the relative motion of the Sun and the local interstellar cloud. This material, mostly hydrogen atoms with about 15\% helium, flows from the Galactic direction (l,b)$\sim$(3$^\circ$,16$^\circ$), placing the Earth downstream of the Sun in early December. The trajectories of the neutral interstellar helium atoms are governed primarily by gravity, executing hyperbolic Keplerian orbits and forming a relatively high-density focusing cone downstream of the Sun about 6$^\circ$ below the ecliptic plane. Interstellar hydrogen, on the other hand, also experiences significant impact from radiation pressure and ionization, creating a neutral hydrogen cavity around the Sun, with negligible focusing.

To maximize the signal from SWCX, {\it DXL} started scanning from $l=140^\circ, b=0^\circ$, to $l=185^\circ, b=-17.5^\circ$ and returned, to scan through the predicted location of the He focusing cone twice (Figure ~\ref{fig: scanpath}). It then performed a fast 360 degree scan of the sky to normalize the data outside the cone and to estimate the instrument background (while looking at the Earth). After the fast scan, the mission again performed two slow scans between $l=185^\circ, b=-17.5^\circ$, and $l=140^\circ, b=0^\circ$ \citep{Galeazzi2014}. Figure~\ref{fig: acs} shows the center of the {\it DXL} scan path as a function of time during launch.

\section{Data Analysis}
\label{analysis}

Both {\it DXL} and {\it ROSAT} measured a combination of SWCX and non-SWCX emission. For the December 2012 DXL observation, the Earth was in the He focusing cone and the line of sight was chosen to pass close the greatest SWCX contribution from the focusing cone, while avoiding bright point sources. The {\it ROSAT} observation in the same direction, taken from the RASS occurred in September of 1990, when the Earth-Sun line was perpendicular to the He focusing cone. The RASS line of sight at that time did not pass through the focusing cone (although it did, like all lines of sight include heliospheric SWCX emission). Thus the {\it DXL} and {\it ROSAT} data contain the same cosmic emission but very different levels of SWCX emission. Given a model of the neutral density distribution in the heliosphere and measurements of the solar wind flux during both observations, one can solve for the true cosmic emission.

The {\it DXL} scan path was designed to run close to the Galactic plane to minimize any non-local contribution to the diffuse X-ray emission. For both {\it DXL} and {\it ROSAT}, the observed flux ($F$) contains contributions from the heliospheric SWCX ($S(t)$) and the non-SWCX (mostly LHB - $L$) emission. The {\it ROSAT} data may also include residual Geocoronal SWCX ($G$), although \cite{kuntz2015} has shown that this residual is small. Any Geocoronal contribution to {\it DXL} is negligible due to the look direction, which is directly away from the Sun. 
We can therefore write the flux for each mission as a function of the pointing direction ($\ell,b$) as:
\begin{equation}
\begin{split}
F_{RASS}(\ell,b,t) = & S_{RASS}(\ell,b,t) + L_{RASS}(\ell,b)+G \\
F_{DXL}(\ell,b,t) = & S_{DXL}(\ell,b,t) + L_{DXL}(\ell,b)
\end{split}
\label{eq:0}
\end{equation}
where the SWCX count rate $S(\ell,b,t)$ is the integral along the line of sight of the product of the ion flux, neutral densities, cross sections for producing charge exchange, and the efficiency summed over all of the lines that fall within the bandpass:

\begin{equation}
\begin{split}
S(\ell,b,t) =& \int\sum_{i}\sum_{j}n_i(t)n_{He}v_{rel}(t)\sigma_ib_{ij}g_jds \\
 & + \int\sum_{i}\sum_{j}n_i(t)n_{H}v_{rel}(t)\sigma_ib_{ij}g_jds,
\end{split}
\label{eq:1}
\end{equation}
where $i$ represents the solar wind species and $j$ the emission lines for the species, $\sigma_i$ is the neutral-dependent 
interaction cross sections for individual species, $b_{ij}$ is the neutral-dependent line branching ratio, $g_j$ is the instruments response to line $j$, and $v_{rel}(t)$ is the relative speed between solar wind and neutral flow (the quadrature sum of bulk and 
thermal velocities).  
With ion density $n_i$ in terms of the proton density $n_p$ at $R_0=1 AU$, and assuming that it scales as one over the distance $R$ from the Sun squared and that solar wind ion neutralizations are minimal, we can define the compound cross-section as \citep{Galeazzi2014} :

\begin{equation}
\alpha=\sum_{i}\sum_{j}\frac{n_i(R_0)}{n_p(R_0)}\sigma_ib_{ij}g_j
\label{eq:2}
\end{equation}

In the case of constant solar-wind conditions, the solar-wind flux can be removed from the integrals, and the total charge exchange rate with H and He can be written as:
\begin{equation}
S(\ell,b,t)=n_p(R_0,t)v_{rel}(t)\alpha_{He}\bigg( \int\frac{n_{He}}{R^2}ds+\frac{\alpha_H}{\alpha_{He}}\int\frac{n_H}{R^2}ds\bigg)
\label{eq:3}
\end{equation}

where $\int\frac{n}{R^2}ds$ is the integrated neutral column density along the line of sight, weighted by one over the distance from the Sun squared $(\frac{1}{R^2})$ to reflect the dilution of the solar wind as it flows outward. Note that in the equation above, the assumed stationarity of the ion flux is an approximation because the flux is known to vary strongly on timescales of about a day. These fluctuations, however, are smoothed when averaging over the few-week transit time through the relevant interplanetary region, as is evident from the very good agreement between sky surveys performed by different missions in different years \citep{Snowden1995a}. We can therefore approximate the time dependent parameters $n_p(R_0,t)$ and $v_{rel}(t)$ as isotropic and ``constant'' in time for the duration of each observation, but assume that they can change by a collective adjustable factor over the 20-year interval between the {\it DXL} and {\it ROSAT} observations.

For simplicity, we can then rewrite Eq.~\ref{eq:3} in terms of two parameters:
\begin{equation}
S(\ell,b,t)=\beta(t) \times N(\ell,b) 
\end{equation}
where
\begin{equation}
\beta(t)=n_p(R_0,t)v_{rel}(t)\alpha_{He} 
\end{equation}
depends on the solar wind properties and the cross section with neutrals, and
\begin{equation}
N(\ell,b)=\int\frac{n_{He}}{R^2}ds+\frac{\alpha_H}{\alpha_{He}}\int\frac{n_H}{R^2}ds
\end{equation}
depends on the pointing direction and the ratio between cross sections with H and He. 

\begin{figure}
\figurenum{5}
\plotone{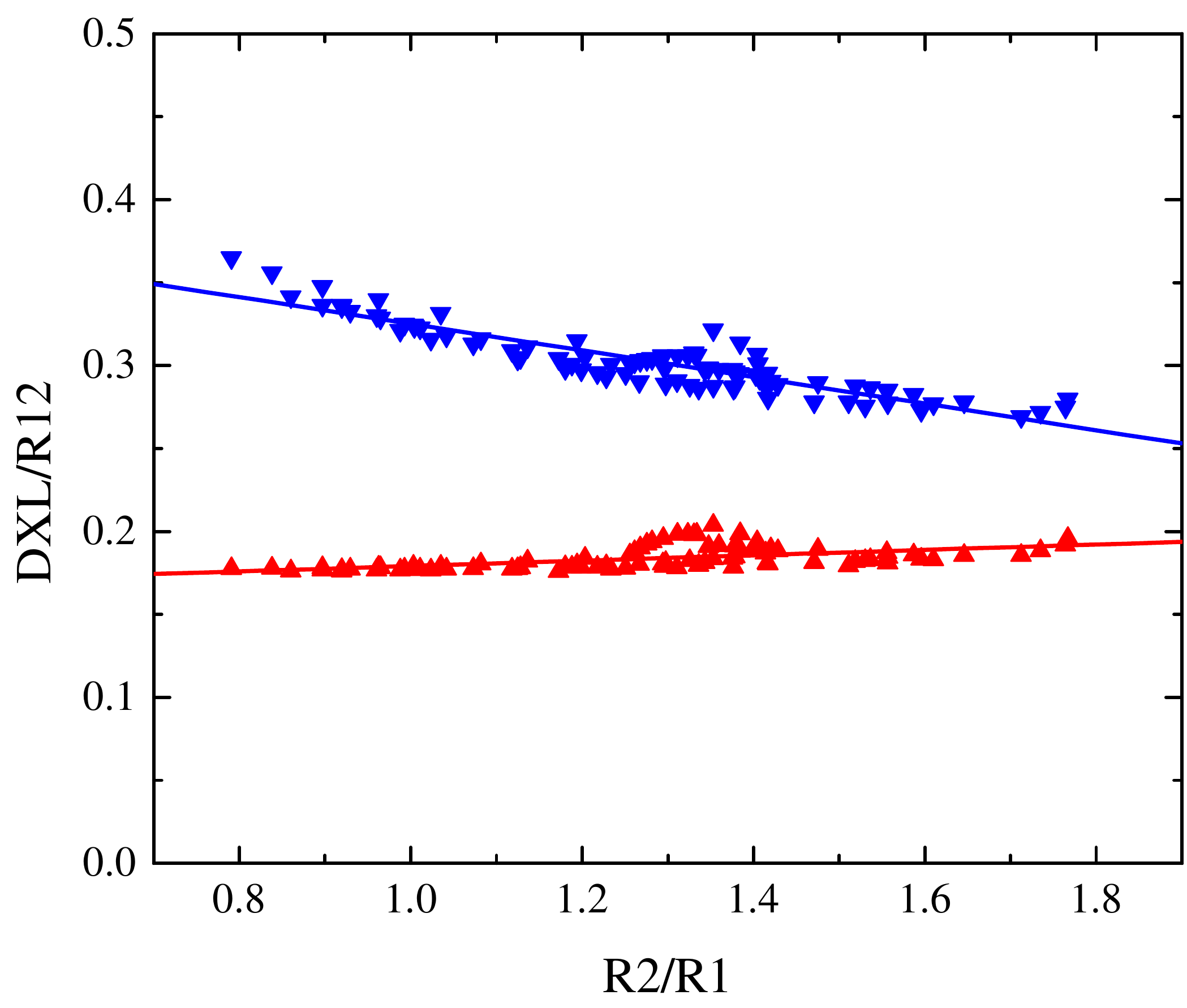}
\caption{{\it DXL/ROSAT} rate vs R2/R1 ratio for the R12 band. The red dots are for counter 1, blue for counter 2. The {\it DXL} rates were calculated by folding various thermal and abundances models for a wide range of
temperatures of the unabsorbed component (to cover the widest range of {\it ROSAT} R2/R1
ratio) through the {\it DXL} response. The temperature and normalization of the absorbed component,
and the photon index and normalization of the power law were fixed to the best fit values
obtained by fitting {\it ROSAT} data at the center of the {\it DXL} slow scan region.}
\label{fig: ratio}
\end{figure}

Direct comparison of {\it DXL} and RASS count rates would require, in general, detailed instrument responses as well as the spectral shape of the emission. However, because the {\it DXL} and RASS responses are very similar, the ratio can be parametrized by a single parameter. For example, the ratio between the {\it DXL} D1+D2 and the RASS R1+R2 count rates is adequately parameterized as a function of R2/R1, almost independently of the underlying spectral model or abundance ratios (see Fig.~\ref{fig: ratio}). The {\it DXL} to RASS count rate ratio shown in Fig.~\ref{fig: ratio} as a function of R2/R1 was calculated from spectral model fits to the RASS data along the {\it DXL} scan path. The model included an unabsorbed thermal component (SWCX+LHB), an absorbed thermal component (Galactic halo), and an absorbed power law (unresolved point sources). We evaluated the neutral hydrogen column density value, required to model the absorption coefficient, using the Infrared Astronomy Satellite (IRAS) 100 $\mu$m data. For the contribution of unresolved point sources, we used a power law with photon index fixed at 1.4 \citep{Masui2009,Yoshino2009}. 
For the unabsorbed component, we tested the APEC \citep{Smith2001}, MEKAL \citep{Mewe85, Mewe86, Liedahl95} and Raymond-Smith \citep{Raymond77} thermal models\footnote{http://www.atomdb.org/}, and \cite{Savage1996} and \cite{Anders1989} abundance models. Once we had determined the best fit, we varied the foreground LHB+SWCX component over a wide range of temperatures and abundances to determine the dependence of the conversion function from {\it ROSAT} to {\it DXL} upon R2/R1 ($D(R2/R1)$, shown in Fig.~\ref{fig: ratio}). Similar procedures have been used for the individual R1 and R2 band, R4 and R5 (using the R5/R4 ratio), and R6 and R7 (using the R7/R6 ratio).

We can then rewrite Eq.~\ref{eq:0} as:
\begin{equation}
F_{RASS}(\ell,b) = \beta_{RASS}(t) \times N_{RASS}(\ell,b) + L_{RASS}(\ell,b)+G
\label{eq:rass}
\end{equation}
\begin{equation}
\begin{split}
F_{DXL}(\ell,b) =& C\times D(R2/R1)\times \\
& (\beta_{DXL}(t) \times N_{DXL}(\ell,b) + L_{RASS}(\ell,b))
\end{split}
\label{eq:flux}
\end{equation}
where $C$ is a constant parameter to account for differences between the laboratory measured response function and the flight response (should be consistent with 1). Notice that in Eq.~\ref{eq:flux} the parameter $D$ is a function of $R2/R1$, $R5/R4$, or $R7/R6$, depending on the band that is being fit. 


\begin{figure}
\figurenum{6}
\plotone{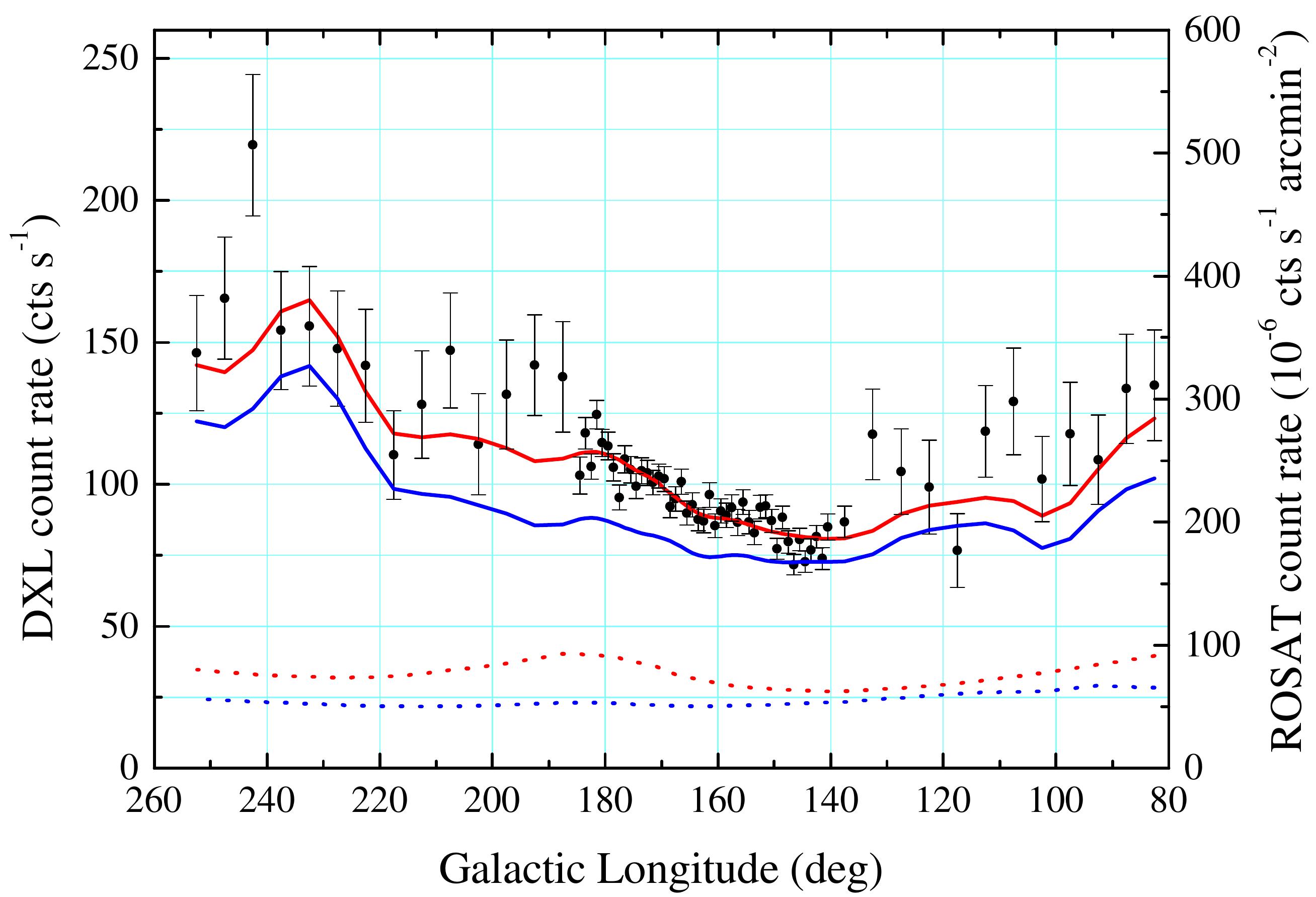}
\caption{{\it DXL} (black dots) count rates in D1 band and {\it ROSAT} (solid blue line) count rates in R1 band. The solid red line shows the best fit to the {\it DXL} count rate. The solar wind charge exchange contributions to {\it DXL} (dashed red line) and {\it ROSAT} (dashed blue line) are also shown.}
\label{fig: D1rate}
\end{figure}

\begin{figure}
\figurenum{7}
\plotone{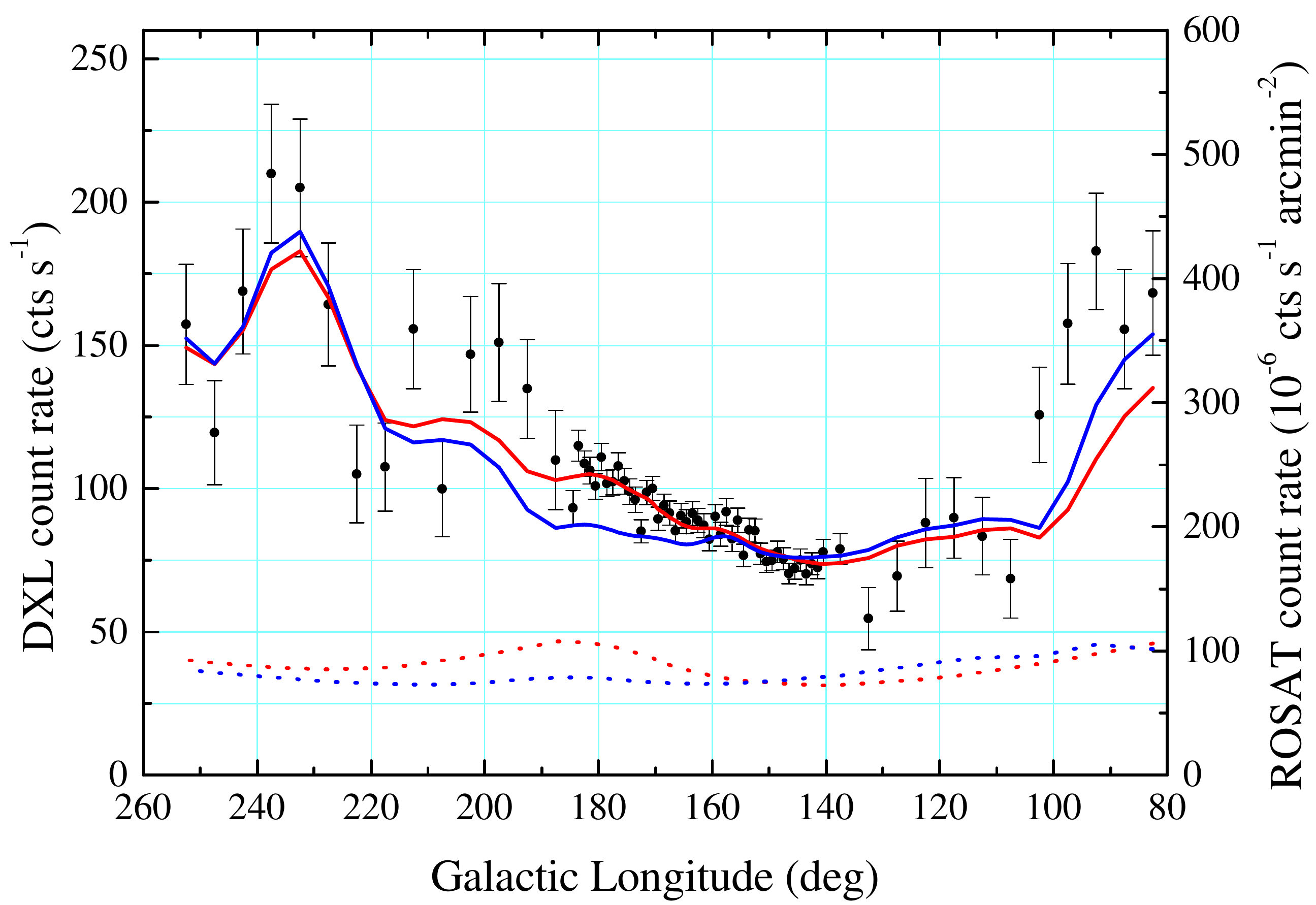}
\caption{Same as Fig.~\ref{fig: D1rate} for the D2 band.}
\label{fig: D2rate}
\end{figure}

\begin{figure}
\figurenum{8}
\plotone{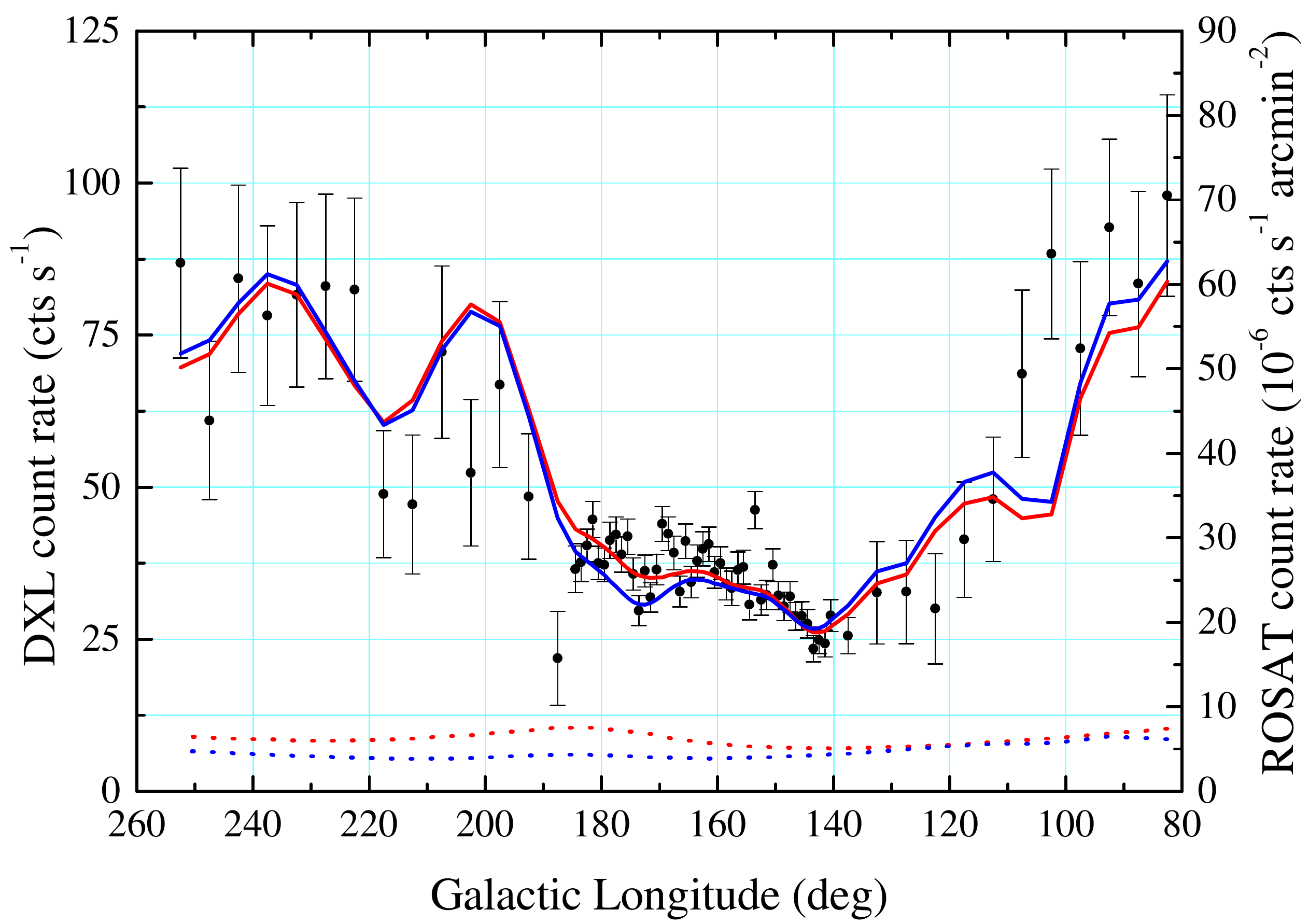}
\caption{Same as Fig.~\ref{fig: D1rate} for the D4 band.}
\label{fig: D4rate}
\end{figure}

\begin{figure}
\figurenum{9}
\plotone{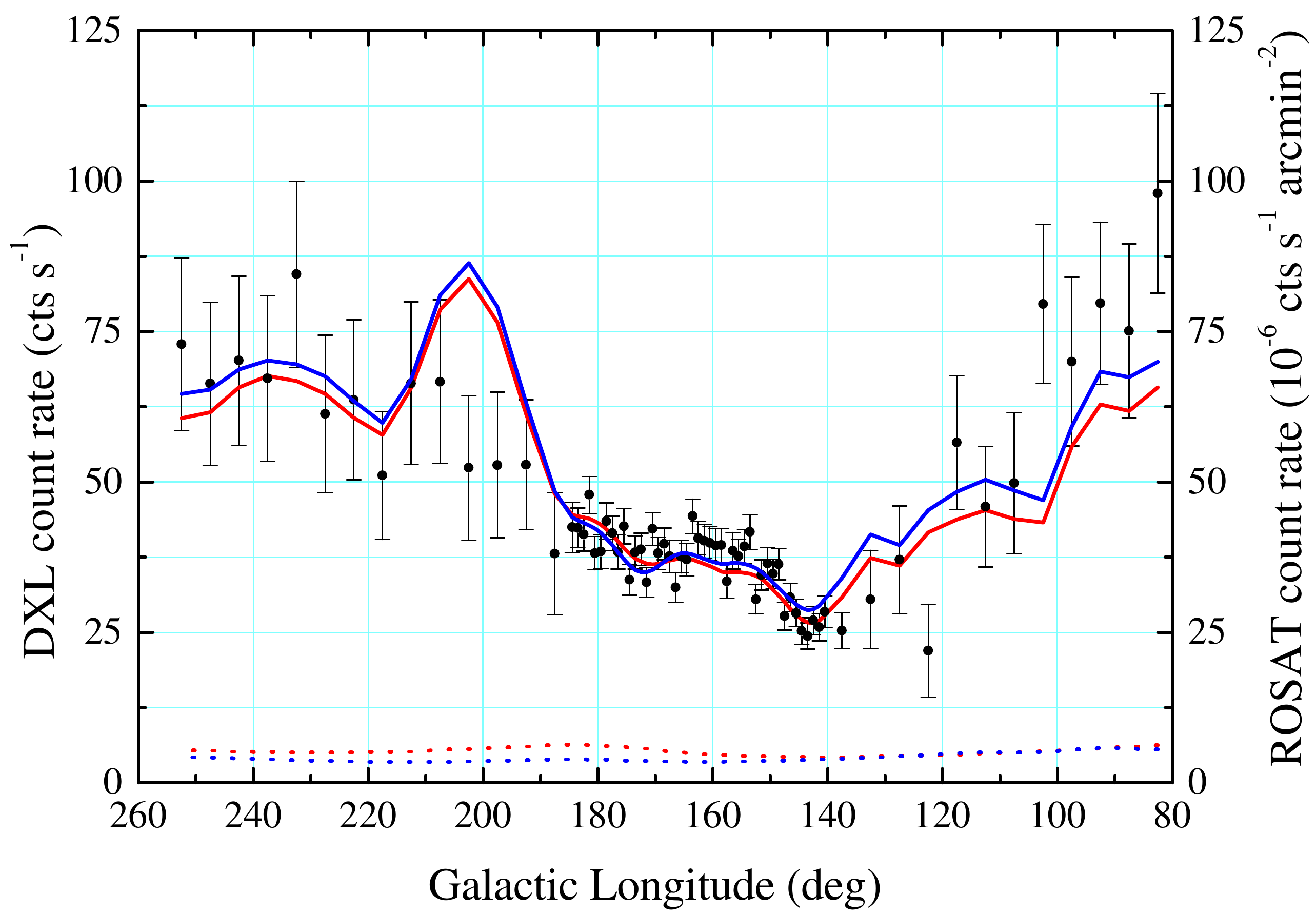}
\caption{Same as Fig.~\ref{fig: D1rate} for the D5 band.}
\label{fig: D5rate}
\end{figure}

For the analysis of the data, we reduced the {\it DXL} data for each band as a function of Galactic longitude (a good proxy for scan path) using 1 degree bins along the slow scan path, and 5 degree bins along the fast scan path. We also extracted {\it ROSAT} data in the six bands for direct comparison with {\it DXL}. The {\it ROSAT} rates were weighted over the {\it DXL} collimators using the same set of longitude bins used for {\it DXL}. Since the {\it RASS} data as published no longer contain the point sources, point source count rates were added to the {\it ROSAT} rates prior to data fitting to account for the contribution of the point sources to the {\it DXL} rates. The point source rates were obtained from the ROSAT bright and faint source catalog in the DXL look direction.\footnote{http://www.xray.mpe.mpg.de/rosat/survey/}. The source rates were weighted by the DXL collimator response, and corrected for vignetting before adding to the RASS band rates. We found that the point sources contribute between 1\% and 5\% to the R12 band, and between 2\% and 15\% to the R45 band along the {\it DXL} scan path.

Following the fitting procedure of \cite{Galeazzi2014} we performed a simultaneous fit to both {\it DXL} counters using Eqs.~\ref{eq:rass} and \ref{eq:flux}. We used, as free parameters, the constants $C_i$ and $C_{ii}$ for each counter, the RASS SWCX parameter $\beta_{RASS}$, and the ratio $r=\frac{\beta_{DXL}}{\beta_{RASS}}$. Note that we decided to use $r$ instead of $\beta_{DXL}$ as it simplifies the fitting procedure and can be more easily interpreted (it is just a constant accounting for the difference in solar wind flux between the time of the {\it DXL} and {\it ROSAT} measurements).  According to solar wind data from NASA's multi-source solar wind database {\it OMNI}\footnote{http://omniweb.gsfc.nasa.gov}, $r$ should be close or slightly smaller than 1, as solar wind flux was somewhat weaker during the {\it DXL} campaign). 

The {\it ROSAT} (solid blue line) count rates for R1, R2, R4 and R5 bands, and {\it DXL} (black dots) count rates for D1, D2, D4 and D5 bands are plotted against Galactic longitude in figures ~\ref{fig: D1rate} through figure ~\ref{fig: D5rate}. The respective figures also show the best fit lines (solid red) to {\it DXL} data, along with SWCX contributions to {\it DXL} (dashed red) and {\it ROSAT} (dashed blue) bands. 

\begin{table*}
\begin{center}
\caption{Summary of the best-fit parameters (described in the text) derived from a fit to {\it DXL} and {\it ROSAT} data in the R12 band. The table highlights the results from some of the different combination of the parameters $\frac{\alpha_H}{\alpha_{He}}$ and $G$.}
\begin{tabular}{cccccccc}
\hline 
\hline
Parameter\footnote{Errors are 1 sigma} &  &  &  &  &  & &  \\ 
\hline 
$\frac{\alpha_H}{\alpha_{He}}$ & 1 & 1   &  2 & 2   & 4 &  6 & 8 \\
$G$                                           & 0  & 50 & 0 & 50 & 0 & 0 & 0 \\ 
$C_i$ & 0.90$\pm$0.06 & 0.94$\pm$0.06 & 0.86$\pm$0.07 & 0.91$\pm$0.07 & 0.80$\pm$0.07 & 0.77$\pm$0.07 & 0.76$\pm$0.07  \\ 
$C_{ii}$ & 0.99$\pm$0.07 & 1.04$\pm$0.07 & 0.95$\pm$0.07 & 1.01$\pm$0.07 & 0.88$\pm$0.08 & 0.85$\pm$0.08 & 0.84$\pm$0.08 \\ 
$\beta_{RASS}=n_p(R_0)v_{rel}\alpha_{He} $(RU cm$^3$ AU) & 5259$\pm$764 & 3052$\pm$721 & 4126$\pm$494 & 2712$\pm$465 & 3418$\pm$423 &  3212$\pm$536 &  3027$\pm$564 \\ 
$r=\frac{\beta_{DXL}}{\beta_{RASS}}$ & 0.44$\pm$ 0.09 & 0.65$\pm$0.19 & 0.53$\pm$0.12 & 0.74$\pm$0.19 & 0.70$\pm$0.14 & 0.77$\pm$0.12 & 0.79$\pm$0.11 \\ 
$\chi^2$(136 d.o.f) & 226 & 215 & 224 & 213 & 221 & 217 & 216 \\ 
On plane SWCX (RU)\footnote{({\it ROSAT} Units = $10^{-6}$ counts s$^{-1}$ arcmin$^{-2}$)} & 124$\pm$18 & 122$\pm$17 & 123$\pm$15 & 131$\pm$14 & 146$\pm$18 & 179$\pm$30 & 208$\pm$39 \\
\hline 
\end{tabular}
\label{tab:a0} 
\end{center}
\end{table*}

\begin{table*}
\begin{center}
\caption{Summary of the best-fit parameters (described in the text) derived from a fit to {\it DXL} and {\it ROSAT} data.}
\begin{tabular}{ccccccc}
\hline 
\hline
Parameter\footnote{Errors are 1 sigma} \footnote{The results shown are calculated using $\frac{\alpha_H}{\alpha_{He}}$=2, $G=50$~RU for the R12 band, $G=25$~RU for the R1 band, $G=25$~RU for the R2 band, and $G=0$~RU for all other bands. } & R1 & R2 & R12 & R4 & R5 & R45 \\ 
\hline 
$C_i$ & 0.84$\pm$0.10 & 0.86$\pm$0.08 & 0.91$\pm$0.07 & 0.94$\pm$0.08 & 1.03$\pm$0.10 & 1.02$\pm$0.06 \\ 
$C_{ii}$ & 0.92$\pm$0.11 & 1.1$\pm$0.11 & 1.01$\pm$0.07 & 1.12$\pm$0.09 & 0.91$\pm$0.08 & 1.03$\pm$0.06 \\ 
$\beta_{RASS}=n_p(R_0)v_{rel}\alpha_{He}$ (RU cm$^3$ AU)\footnote{Using data from the WIND satellite we calculated the DXL solar wind proton flux $n_p(R_o)v_{rel}=2.7\times 10^8 cm^{-2} s^{-1}$.} & 1040$\pm$352 & 1810$\pm$341 & 2712$\pm$465 & 123$\pm$79 & 144$\pm$113 & 218$\pm$131 \\ 
$r=\frac{\beta_{DXL}}{\beta_{RASS}}$ & 1.00$\pm$0.48 & 0.76$\pm$0.20 & 0.74$\pm$0.19 & 1.00$\pm$0.51 & 1.00$\pm$0.64 & 1.00$\pm$0.48 \\ 
$\chi^2$(136 d.o.f) & 184 & 181 & 213 & 238 & 221 & 359 \\ 
\hline 
\end{tabular}
\label{tab:a} 
\end{center}
\end{table*}

\begin{table*}
\begin{center}
\caption{SWCX Contributions to {\it ROSAT} bands with statistical and systematic errors respectively.}
\begin{tabular}{ccccccc}
\hline
\hline
SWCX\footnote{Statistical errors are 1 sigma, systematic errors are absolute.} \footnote{The results shown are calculated using $\frac{\alpha_H}{\alpha_{He}}$=2, $G=50$~RU for the R12 band, $G=25$~RU for the R1 band, $G=25$~RU for the R2 band, and $G=0$~RU for all other bands. } & R1 & R2 & R12\footnote{\label{note}The R12 and R45 results come from independent fits of the data in the two combined bands and are consistent with the sum of the SWCX contribution from the individual bands.} & R4 & R5 & R45\textsuperscript{\ref{note}} \\
\hline
$l\sim140.5^\circ, b\sim0^\circ)$ (RU) & 56$\pm$11$\pm$20 & 79$\pm$10$\pm$7 & 131$\pm$14$\pm$30  & 4$\pm$2$\pm$2 & 4$\pm$4$\pm$3 & 7$\pm$4$\pm$4\\
$l\sim140.5^\circ, b\sim0^\circ)$ (\%) & 33$\pm$6$\pm$12 & 44$\pm$6$\pm$5 & 38$\pm$4$\pm$8  & 18$\pm$12$\pm$11 & 14$\pm$11$\pm$9 & 13$\pm$8$\pm$8 \\
\hline
\end{tabular}
\label{tab:b}
\end{center}
\end{table*}

\begin{table*}
\begin{center}
\caption{Extrapolated SWCX contributions to {\it ROSAT} bands averaged across the whole sky.}
\begin{tabular}{ccccccc}
\hline
\hline
SWCX\footnote{Statistical errors are 1 sigma, systematic errors are absolute.} \footnote{The results shown are calculated using $\frac{\alpha_H}{\alpha_{He}}$=2, $G=50$~RU for the R12 band, $G=25$~RU for the R1 band, $G=25$~RU for the R2 band, and $G=0$~RU for all other bands. }\footnote{\label{note}The all-sky average were derived from the values of $\beta_{RASS}$ and R from Table~\ref{tab:a} and the model of the neutral distribution from \cite{Koutroumpa2006} } & R1 & R2 & R12 & R4 & R5 & R45 \\
\hline
All-sky average (RU) & 70$\pm$15$\pm$33 & 103$\pm$15$\pm$14 & 168$\pm$20$\pm$53 & 5$\pm$3$\pm$4 & 6$\pm$5$\pm$5 & 9$\pm$6$\pm$7 \\
All-sky average (\%) & 26$\pm$6$\pm$13 & 30$\pm$4$\pm$4 & 27$\pm$3$\pm$9 & 8$\pm$5$\pm$5 & 7$\pm$6$\pm$5 & 6$\pm$4$\pm$4 \\
\hline
\end{tabular}
\label{tab:c}
\end{center}
\end{table*}

\section{Results}
\label{Results}
\subsection{\it DXL Results}

The result for the combined RASS R12 band has already been reported and discussed in \cite{Galeazzi2014}. Here 
we report our results on all the RASS individual bands and their implications. 
In \cite{Galeazzi2014}, two possible contributions from geocoronal SWCX and two different ratios of H to He compound 
cross-sections were used. 
For this work, we also tested different values of the contribution $G$ from geocoronal SWCX to the 
individual R1 and R2 bands, and we tested different values of the H to He compound cross-section ratio $\frac{\alpha_H}{\alpha_{He}}$ from 1 to 8, as there 
is theoretical indication that the ratio may be higher than the maximum value of 2 considered in the previous analysis. (Vasili Kharchenko, private communication). The comparison between \cite{Galeazzi2014} and \cite{kuntz2015} also indicates that the ratio may be higher.
The fit result depends weakly on the ratio $\frac{\alpha_H}{\alpha_{He}}$, preventing us from successfully fitting independently for $\alpha_H$ and $\alpha_{He}$. For example, in the R12 band, the SWCX contribution on the Galactic plane slowly changes by less than a factor of 2 while the ratio is increased by a factor of 8. Table~\ref{tab:a0} shows the individual parameters for some of the combinations in R12 band. Similar or weaker trends have been found in all other bands. As a consequence, from here on, we show one nominal set of values and include the variation due to different values of $G$ and $\frac{\alpha_H}{\alpha_{He}}$ in the systematic error (up to a ratio of 4). The systematic error also includes uncertainties in the laboratory calibration (represented by $C_i$ and $C_{ii}$), differences in the response function between {\it ROSAT} and {\it DXL} (represented by $D(R2/R1)$), and the effect of the data range used for the fits (for example, using only the slow scan region vs. the full scan, half the data, etc.).

As expected, the SWCX contribution to the RASS R6 and R7 bands was found to be compatible with zero, therefore, for the rest of this paper, we focus only on the R1-R5 bands. 

The fitting parameters obtained in all the bands are presented in 
Table~\ref{tab:a}. The table shows the representative data for the condition when the ratio of H to He cross-sections was 2 and the geocoronal contribution to the RASS was assumed to be $50$~RU ({\it ROSAT} Units or $10^{-6}$~counts~s$^{-1}$~arcmin$^{-2}$) for the R12 band - evenly split between the two bands - and 
0~RU for the R45 band. As we previously discussed, changes in the results due to different assumptions are included in the systematic error (up to a ratio $\frac{\alpha_H}{\alpha_{He}}$=4).  Following the same convention used in \cite{Galeazzi2014}, the parameters $C_i$ and $C_{ii}$ are
the ratios of the fitted DXL response to the nominal value from laboratory conditions for the two {\it DXL} counters. The errors in the table are 1 $\sigma$. 
We note that the values of $\chi ^2$ are larger than optimal. The {\it DXL} mission was designed to focus on the slow scan region, as it is the region that constrains the SWCX to LHB ratio near the Galactic plane. In that region the reduced $\chi ^2$ is well within statistically acceptable values for all fits. However, outside the slow scan region, particularly at very low longitude, there are additional effects not included in the model that affect the value of $\chi ^2$, but do not affect the result of the investigation (this was verified by using different fitting ranges).

\begin{figure*}
\figurenum{10}
\plotone{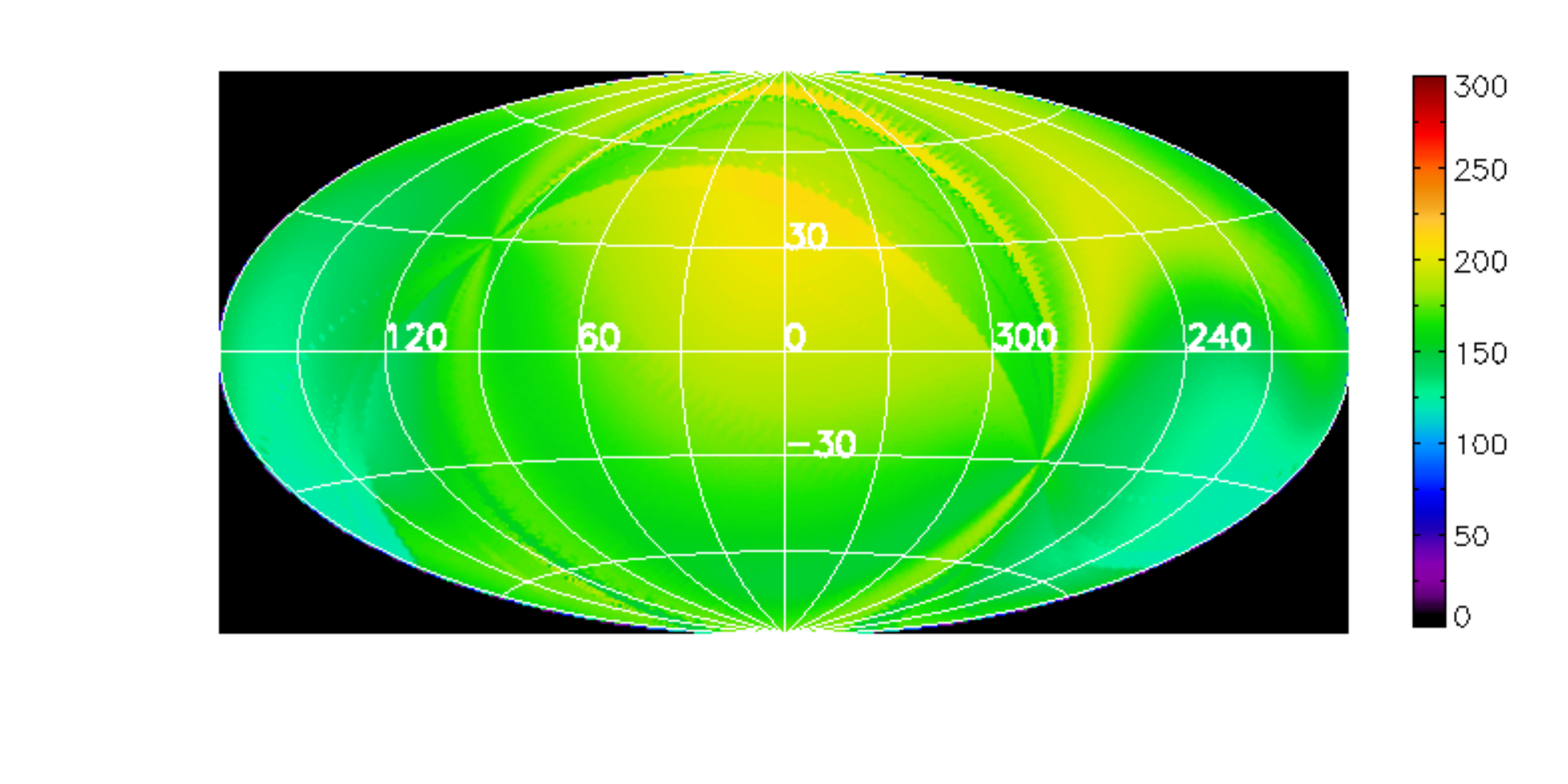}
\caption{The Aitoff-Hammer projection of SWCX contribution to the {\it ROSAT} R12 band. The intensity scale shows the SWCX count rate in RU. Note that the sharp edges visible in this map are due to abrupt shifts in vantage point around the Earth's orbit during the {\it ROSAT}  survey, since the survey comes back to its starting point after six months and there is a missed section that was filled in at a later time.}
\label{fig: R12SWCX}
\figurenum{11}
\plotone{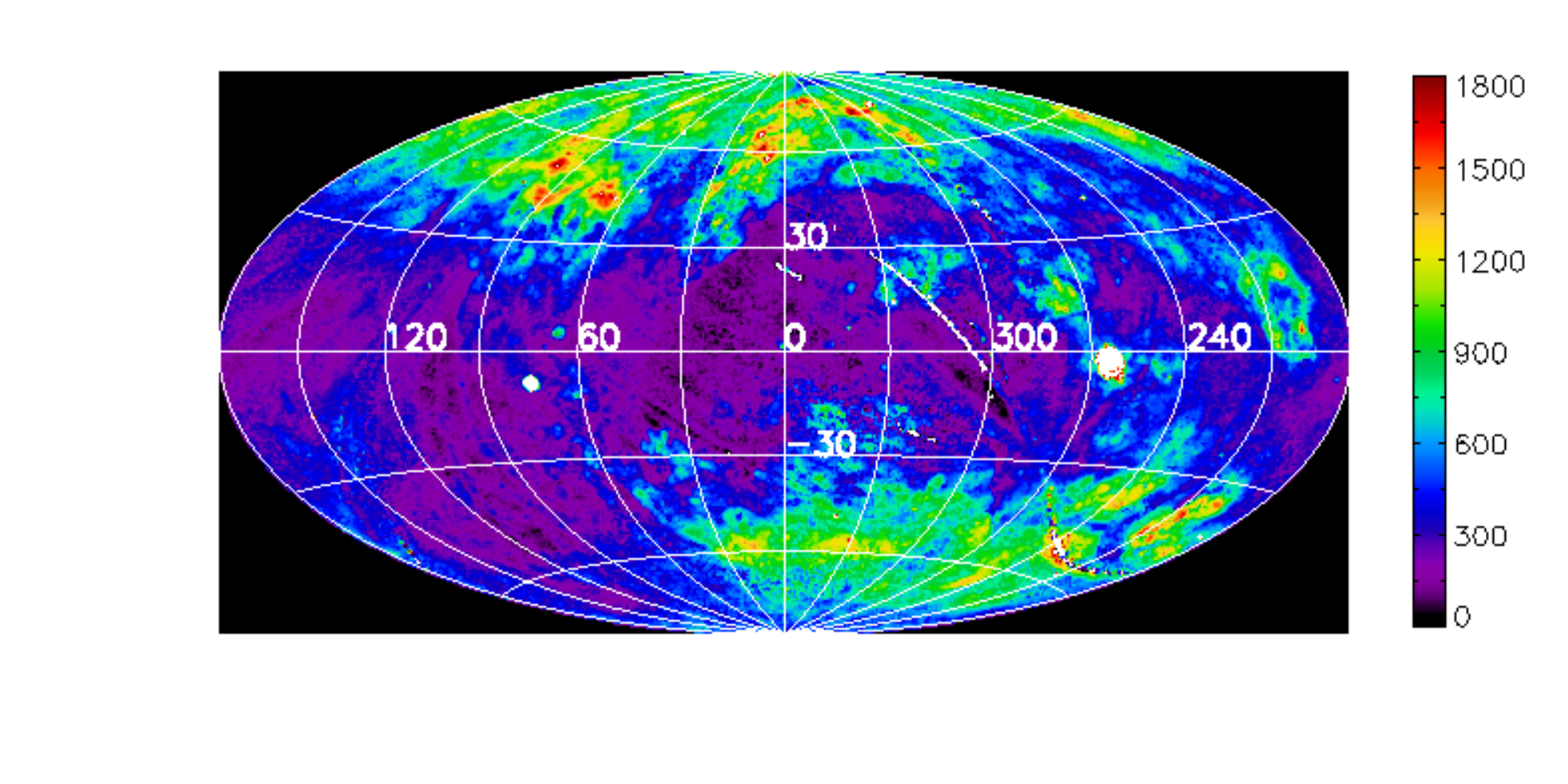}
\caption{The Aitoff-Hammer projection of {\it ROSAT} R12 band in RU after removing the SWCX contribution.}
\label{fig: NoSWCXR12}
\end{figure*}

\begin{figure*}
\figurenum{12}
\plotone{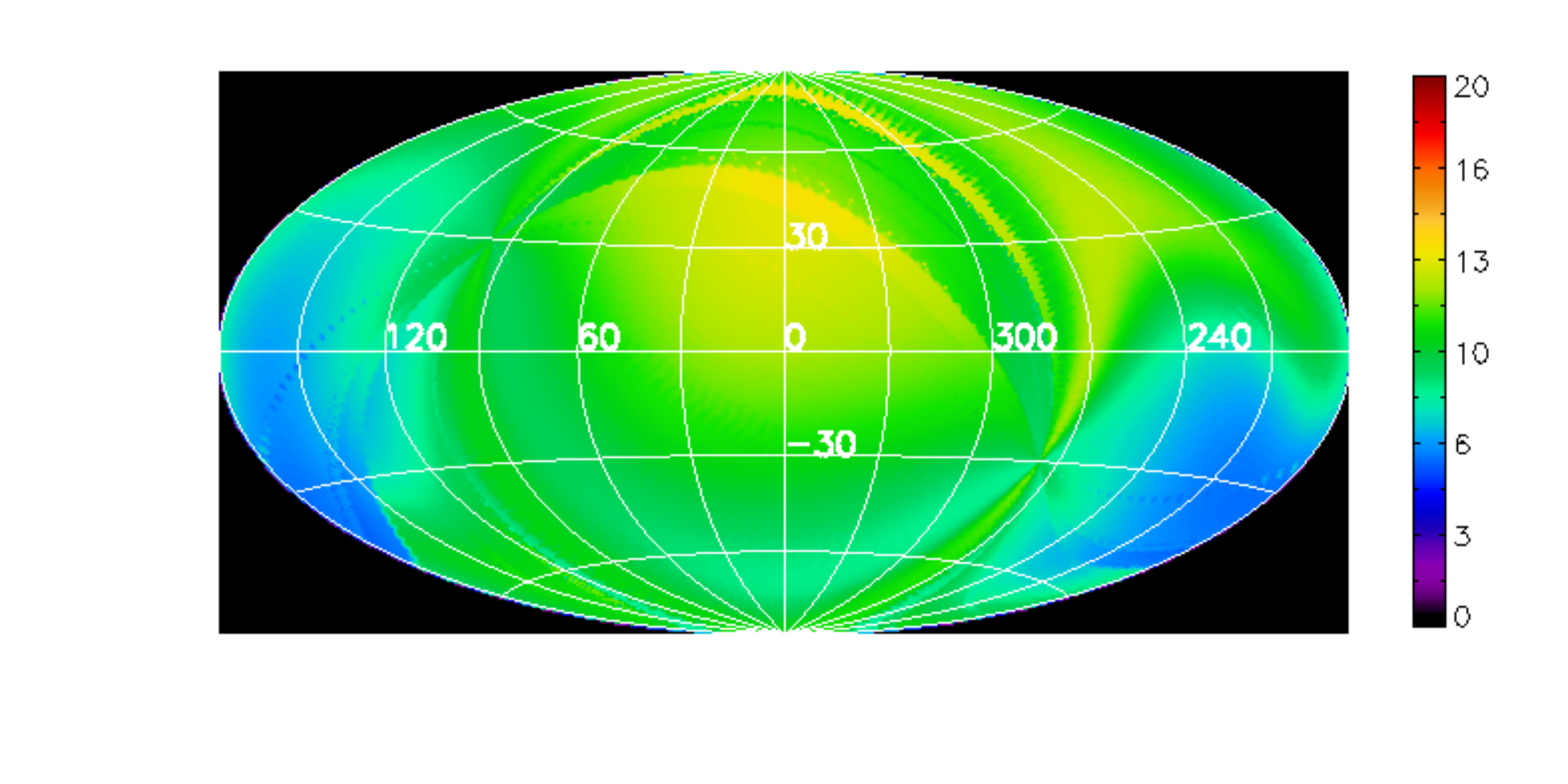}
\caption{The Aitoff-Hammer projection off SWCX contribution in the {\it ROSAT} R45 band. The intensity scale is in {\it ROSAT} Units (RU).}
\label{fig: R45SWCX}
\figurenum{13}
\plotone{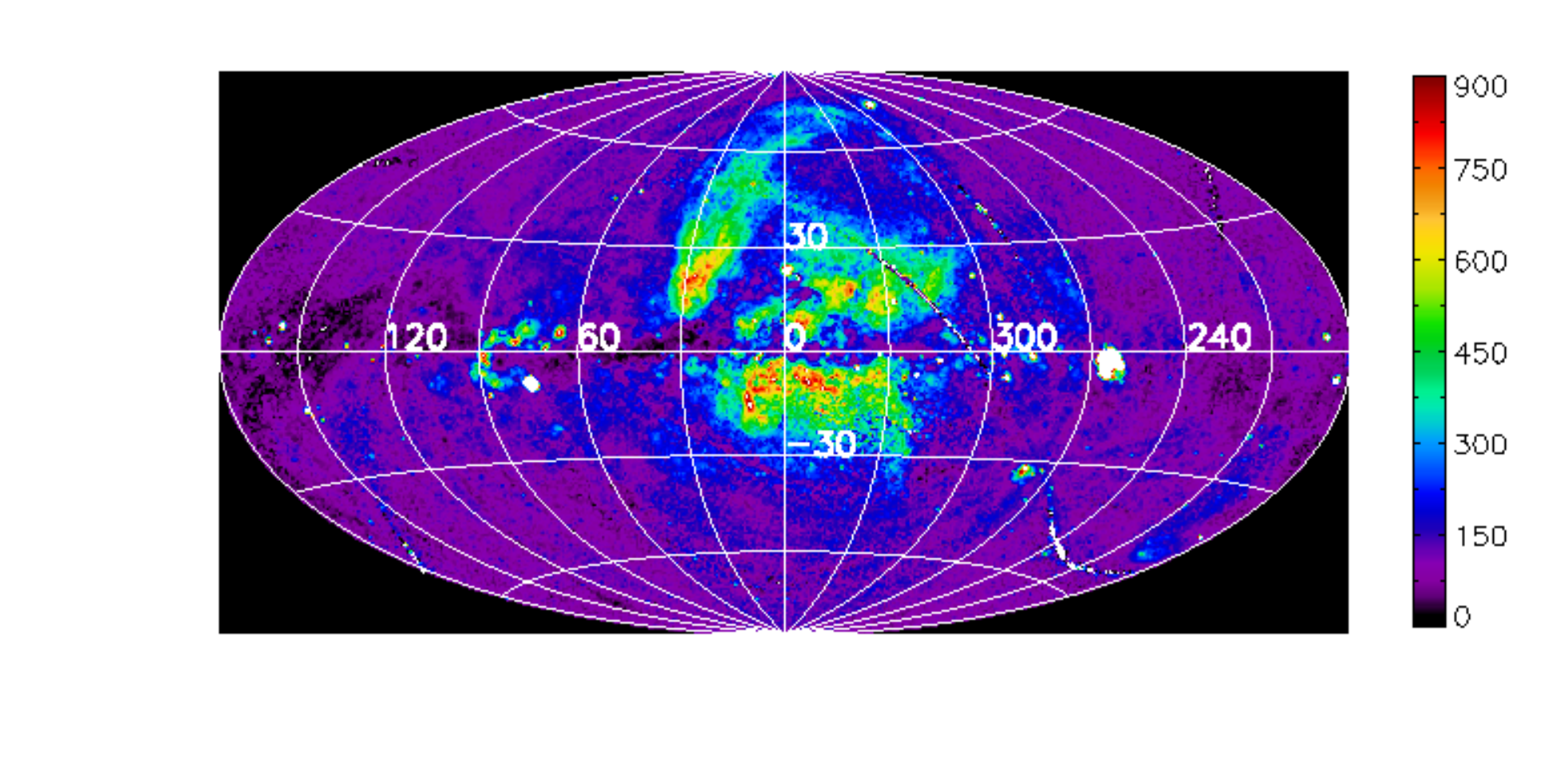}
\caption{ROSAT R45 band map after removing SWCX.}
\label{fig: NoSWCXR45}
\end{figure*}

To better understand the contribution of SWCX to the RASS bands and, more in general, to the diffuse X-ray emission, 
we use the parameters from Table~\ref{tab:a} to calculate both absolute (in RU) and fractional SWCX rates. In particular, 
in Table~\ref{tab:b} we report the SWCX contribution at $l\sim 140.5^o$, $b\sim 0$, which is were the {\it DXL} scan path 
intersects the Galactic plane.  We point out that, according the the ROSAT maps, this is a point where the total diffuse X-ray emission is close to the minimum across the whole sky, and where we therefore expect the highest relative SWCX contribution. Although we used an independent (and slightly different) procedure to reduce the data, the results in the combined 
R12 band are consistent with those already reported in \cite{Galeazzi2014}. In the R12 band, we find that the total SWCX contribution is $131 \pm14 (statistical)\pm30(systematic)$~RU at the Galactic plane. The combined SWCX contribution is $38\%\pm4\%\pm8\%$ of the {\it ROSAT} count rate at the location.

\subsection{Extrapolating the {\it DXL} Results}

We also used the expected distribution of neutral H and He from 
\cite{Koutroumpa2006} to estimate the SWCX contribution over the whole sky (Table~\ref{tab:c}). 
The composite emission coefficients fit in the area of the DXL scan were used for all directions and for the six month duration of the survey, placing considerably more stress on the assumptions of isotropy and constancy in time of the Solar wind.  The systematic uncertainties given for the all-sky averages do not include these additional factors, but since the Solar wind is expected to be most isotropic at Solar maximum when the survey was performed, we estimate that these are small compared to those from the poorly constrained ratio of $\alpha_H/\alpha_{He}$.  The model for the 3-D distribution of neutral H and He over the Solar system has been tested by Solar backscatter and direct in-situ measurements and should not be a major contributor to the uncertainties.
Using this method, the Aitoff-Hammer projection of SWCX contribution in the RASS R12 band is presented in Figure~\ref{fig: R12SWCX}. The estimated all-sky average emission from SWCX in the R12 band is $168\pm20\pm53$~RU, which is $27\%\pm3\%\pm9\%$ of the total R12 emission. We have also used the estimated SWCX emission of Figure~\ref{fig: R12SWCX} to remove its contribution from RASS. The ``cleaned" {\it ROSAT} R12 map is shown in Figure ~\ref{fig: NoSWCXR12}.

Similarly, in the R45 band, the SWCX contributions at $l\sim 140.5^o$, $b\sim 0$ is $7\pm4\pm4$~RU, or $13 \%\pm8\%\pm8 \%$ of the observed {\it ROSAT} counts. As most charge exchange models indicate that the SWCX contribution to the R45 band is dominated by oxygen emission (mostly OVII), we also used this result to get upper limits on the SWCX OVII+OVIII emissions. Specifically, if we assume that all the R45 SWCX emission is due to OVII+OVIII, we can set an upper limit on the emission of $3.2\pm1.7$ LU. This is consistent with predictions from \citet{Koutroumpa2009a} and with observations in the shadows of nearby molecular clouds \citep{Gupta2009b,Henley2013} showing that the small residual contribution to the R45 band from less than the $\sim$150~pc distances to these clouds is dominated by OVII line emission.  Both the apparent variability of these lines in multiple observations of the same cloud and the very small OVII fluxes predicted by the best emission model fits to the LHB suggest that this is largely SWCX emission. Using the average temperature of 0.0903~keV for the LHB, and an emission measure of $2.55\times10^{-3}$~cm$^{-6}$~pc from Liu et. al. (in preparation - see also \cite{Snowden2014}), we calculated that the LHB contributes about 0.37~LU to OVII+OVIII, or $\sim$3~RU to the R45 on the Galactic plane, which is $\sim$4\% of the average {\it ROSAT} rate on the plane. The bulk of the R45 flux must come from non-local emission, identified with Galactic Halo or non-Galactic emission at high latitude. The origin of the bulk of R45 emission on the Galactic plane (where Galactic Halo and non-Galactic emission are absorbed) remains an open question. It seems that an additional, not yet identified source of X-rays must be responsible for about 90\% of the total emission on the Galactic plane in the R45 band. 
This is a long-standing issue in X-ray astronomy, pre-dating the discovery of SWCX \citep{McCammon1990}. 

The Aitoff-Hammer projection of the SWCX contribution to the RASS R45 band is presented in Figure~\ref{fig: R45SWCX}.
We updated the {\it ROSAT} R45 map by removing the SWCX emission as shown in Fig.~\ref{fig: NoSWCXR45}. Notice that the contribution from SWCX is very small, and its effect on the RASS map is barely noticeable. When extrapolated over the whole sky, the average SWCX contribution is $9\pm6\pm7$~RU, or $6 \%\pm4 \%\pm4 \%$ of the total observed rate. 

We also used the ``AtomDB Charge Exchange (ACX)'' model \citep{Smith2014a} to compare our result with the analysis performed by \citet{Smith2014a} using data from the $DXS$ mission \citep{Sanders2001}. Specifically, we performed a global fit to our estimated all-sky SWCX contribution to the R1-R5 bands with ``solar wind temperature'' (i.e., the temperature of the solar corona responsible for the solar wind) and normalization as free parameters. We find a solar wind temperature of $1.22\pm0.08\times 10^{6}$~K, within the range of temperatures obtained by \cite{Smith2014a}.
However, we note that \cite{Smith2014a} find that SWCX is the main contributor to the total 0.1-0.4 keV flux, with the LHB contributing only 26\%$\pm$4\%, which is inconsistent with the results reported here and in \cite{Galeazzi2014}. \cite{Smith2014a} relied on an extremely approximate model of the charge exchange spectrum (due to the lack of good atomic data), and errors in the underlying model could have caused the discrepancy.

\section{Conclusions}
\label{conclusion}

Using data from the {\it DXL} sounding rocket mission, we have made quantitative estimates of the SWCX contribution to the RASS maps of the diffuse X-ray background. This contribution was removed from the RASS maps to produce ``clean'' maps
of the interstellar and extragalactic diffuse emission. 
Averaged over the sky, we found about 27\% of the observed flux in R12 and 6\% in R45 is due to SWCX.  Even at a point in the Galactic plane where the observed fluxes are near their absolute minimum values, SWCX accounts for $\sim$38\% of the R12 band and $\sim$13\% of R45.

\acknowledgments
This work was supported by NASA award numbers NNX11AF04G and NNX09AF09G. We would like to thank the sounding rocket staff at NASA's Wallops Flight Facility and the White Sands Missile Range for their support, technical personnel and undergraduate students at the University of Miami, NASA's Goddard Space Flight Center and the University of Michigan for their support of the instrument's development, and Mark Mulligan at the University of Wisconsin Space Science and Engineering Center (SSEC) for his support with the DXS windows. D.K. and R.L. acknowledge financial support for their activity through the programme `Soleil H\'{e}liosph\'{e}re Magn\'{e}tosph\'{e}re' of the French space agency CNES, and the National Program `Physique Chimie du Milieu Interstellaire' of the Institut National des Sciences de l'Univers (INSU).

\bibliographystyle{apj}

\begin{thebibliography}{}
\expandafter\ifx\csname natexlab\endcsname\relax\def\natexlab#1{#1}\fi

\bibitem[{{Anders} \& {Grevesse}(1989)}]{Anders1989}
{Anders}, E., \& {Grevesse}, N. 1989, \gca, 53, 197

\bibitem[{{Bellm} \& {Vaillancourt}(2005)}]{BellmVaillancourt2005}
{Bellm}, E.~C., \& {Vaillancourt}, J.~E. 2005, \apj, 622, 959

\bibitem[{{Collier} {et~al.}(2015){Collier}, {Porter}, {Sibeck}, {Carter},
  {Chiao}, {Chornay}, {Cravens}, {Galeazzi}, {Keller}, {Koutroumpa},
  {Kujawski}, {Kuntz}, {Read}, {Robertson}, {Sembay}, {Snowden}, {Thomas},
  {Uprety}, \& {Walsh}}]{Collier2015}
{Collier}, M.~R., {Porter}, F.~S., {Sibeck}, D.~G., {et~al.} 2015, Review of
  Scientific Instruments, 86, 071301

\bibitem[{{Cox}(1998)}]{Cox1998}
{Cox}, D.~P. 1998, in Lecture Notes in Physics, Berlin Springer Verlag, Vol.
  506, IAU Colloq. 166: The Local Bubble and Beyond, ed. D.~{Breitschwerdt},
  M.~J. {Freyberg}, \& J.~{Truemper}, 121--131

\bibitem[{{Cravens}(1997)}]{cravens1997}
{Cravens}, T.~E. 1997, \grl, 24, 105

\bibitem[{{Cravens}(2000)}]{Cravens2000}
---. 2000, \apjl, 532, L153

\bibitem[{{Cravens} {et~al.}(2001){Cravens}, {Robertson}, \&
  {Snowden}}]{cravens2001}
{Cravens}, T.~E., {Robertson}, I.~P., \& {Snowden}, S.~L. 2001, \jgr, 106,
  24883

\bibitem[{{Galeazzi} {et~al.}(2011){Galeazzi}, {Chiao}, {Collier}, {Cravens},
  {Koutroumpa}, {Kuntz}, {Lepri}, {McCammon}, {Porter}, {Prasai}, {Robertson},
  {Snowden}, \& {Uprety}}]{Galeazzi2011}
{Galeazzi}, M., {Chiao}, M., {Collier}, M.~R., {et~al.} 2011, Experimental
  Astronomy, 32, 83

\bibitem[{{Galeazzi} {et~al.}(2012){Galeazzi}, {Collier}, {Cravens},
  {Koutroumpa}, {Kuntz}, {Lepri}, {McCammon}, {Porter}, {Prasai}, {Robertson},
  {Snowden}, {Thomas}, \& {Uprety}}]{Galeazzi2012a}
{Galeazzi}, M., {Collier}, M.~R., {Cravens}, T., {et~al.} 2012, Astronomische
  Nachrichten, 333, 383

\bibitem[{{Galeazzi} {et~al.}(2014){Galeazzi}, {Chiao}, {Collier}, {Cravens},
  {Koutroumpa}, {Kuntz}, {Lallement}, {Lepri}, {McCammon}, {Morgan}, {Porter},
  {Robertson}, {Snowden}, {Thomas}, {Uprety}, {Ursino}, \&
  {Walsh}}]{Galeazzi2014}
{Galeazzi}, M., {Chiao}, M., {Collier}, M.~R., {et~al.} 2014, \nat, 512, 171

\bibitem[{{Gupta} {et~al.}(2009){Gupta}, {Galeazzi}, {Koutroumpa}, {Smith}, \&
  {Lallement}}]{Gupta2009b}
{Gupta}, A., {Galeazzi}, M., {Koutroumpa}, D., {Smith}, R., \& {Lallement}, R.
  2009, \apj, 707, 644

\bibitem[{{Henley} \& {Shelton}(2008)}]{Henley2008a}
{Henley}, D.~B., \& {Shelton}, R.~L. 2008, \apj, 676, 335

\bibitem[{{Henley} \& {Shelton}(2013)}]{Henley2013}
---. 2013, \apj, 773, 92

\bibitem[{{Koutroumpa} {et~al.}(2009{\natexlab{a}}){Koutroumpa}, {Collier},
  {Kuntz}, {Lallement}, \& {Snowden}}]{Koutroumpa2009a}
{Koutroumpa}, D., {Collier}, M.~R., {Kuntz}, K.~D., {Lallement}, R., \&
  {Snowden}, S.~L. 2009{\natexlab{a}}, \apj, 697, 1214

\bibitem[{{Koutroumpa} {et~al.}(2006){Koutroumpa}, {Lallement}, {Kharchenko},
  {Dalgarno}, {Pepino}, {Izmodenov}, \& {Qu{\'e}merais}}]{Koutroumpa2006}
{Koutroumpa}, D., {Lallement}, R., {Kharchenko}, V., {et~al.} 2006, \aap, 460,
  289

\bibitem[{{Koutroumpa} {et~al.}(2009{\natexlab{b}}){Koutroumpa}, {Lallement},
  {Raymond}, \& {Kharchenko}}]{Koutroumpa2009b}
{Koutroumpa}, D., {Lallement}, R., {Raymond}, J.~C., \& {Kharchenko}, V.
  2009{\natexlab{b}}, \apj, 696, 1517

\bibitem[{{Kuntz} \& {Snowden}(2000)}]{KuntzSnowden2000}
{Kuntz}, K.~D., \& {Snowden}, S.~L. 2000, \apj, 543, 195

\bibitem[{{Kuntz} {et~al.}(2015){Kuntz}, {Collado-Vega}, {Collier}, {Connor},
  {Cravens}, {Koutroumpa}, {Porter}, {Robertson}, {Sibeck}, {Snowden},
  {Thomas}, \& {Walsh}}]{kuntz2015}
{Kuntz}, K.~D., {Collado-Vega}, Y.~M., {Collier}, M.~R., {et~al.} 2015, \apj,
  808, 143

\bibitem[{{Lallement} {et~al.}(1985{\natexlab{a}}){Lallement}, {Bertaux}, \&
  {Dalaudier}}]{Lallement1985a}
{Lallement}, R., {Bertaux}, J.~L., \& {Dalaudier}, F. 1985{\natexlab{a}}, \aap,
  150, 21

\bibitem[{{Lallement} {et~al.}(1985{\natexlab{b}}){Lallement}, {Bertaux}, \&
  {Kurt}}]{Lallement1985b}
{Lallement}, R., {Bertaux}, J.~L., \& {Kurt}, V.~G. 1985{\natexlab{b}}, \jgr,
  90, 1413

\bibitem[{{Lallement} {et~al.}(2004){Lallement}, {Raymond}, {Vallerga},
  {Lemoine}, {Dalaudier}, \& {Bertaux}}]{Lallement2004}
{Lallement}, R., {Raymond}, J.~C., {Vallerga}, J., {et~al.} 2004, \aap, 426,
  875

\bibitem[{{Liedahl} {et~al.}(1995){Liedahl}, {Osterheld}, \&
  {Goldstein}}]{Liedahl95}
{Liedahl}, D.~A., {Osterheld}, A.~L., \& {Goldstein}, W.~H. 1995, \apjl, 438,
  L115

\bibitem[{{Lisse} {et~al.}(1996){Lisse}, {Dennerl}, {Englhauser}, {Harden},
  {Marshall}, {Mumma}, {Petre}, {Pye}, {Ricketts}, {Schmitt}, {Trumper}, \&
  {West}}]{lisse96}
{Lisse}, C.~M., {Dennerl}, K., {Englhauser}, J., {et~al.} 1996, Science, 274,
  205

\bibitem[{{Masui} {et~al.}(2009){Masui}, {Mitsuda}, {Yamasaki}, {Takei},
  {Kimura}, {Yoshino}, \& {McCammon}}]{Masui2009}
{Masui}, K., {Mitsuda}, K., {Yamasaki}, N.~Y., {et~al.} 2009, \pasj, 61, S115

\bibitem[{{McCammon} {et~al.}(1983){McCammon}, {Burrows}, {Sanders}, \&
  {Kraushaar}}]{McCammon1983}
{McCammon}, D., {Burrows}, D.~N., {Sanders}, W.~T., \& {Kraushaar}, W.~L. 1983,
  \apj, 269, 107

\bibitem[{{McCammon} \& {Sanders}(1990)}]{McCammon1990}
{McCammon}, D., \& {Sanders}, W.~T. 1990, \araa, 28, 657

\bibitem[{{Mewe} {et~al.}(1985){Mewe}, {Gronenschild}, \& {van den
  Oord}}]{Mewe85}
{Mewe}, R., {Gronenschild}, E.~H.~B.~M., \& {van den Oord}, G.~H.~J. 1985,
  \aaps, 62, 197

\bibitem[{{Mewe} {et~al.}(1986){Mewe}, {Lemen}, \& {van den Oord}}]{Mewe86}
{Mewe}, R., {Lemen}, J.~R., \& {van den Oord}, G.~H.~J. 1986, \aaps, 65, 511

\bibitem[{{Michels} {et~al.}(2002){Michels}, {Raymond}, {Bertaux},
  {Qu{\'e}merais}, {Lallement}, {Ko}, {Spadaro}, {Gardner}, {Giordano},
  {O'Neal}, {Fineschi}, {Kohl}, {Benna}, {Ciaravella}, {Romoli}, \&
  {Judge}}]{Michels2002}
{Michels}, J.~G., {Raymond}, J.~C., {Bertaux}, J.~L., {et~al.} 2002, \apj, 568,
  385

\bibitem[{{Mushotzky} {et~al.}(2000){Mushotzky}, {Cowie}, {Barger}, \&
  {Arnaud}}]{Mushotzky2000}
{Mushotzky}, R.~F., {Cowie}, L.~L., {Barger}, A.~J., \& {Arnaud}, K.~A. 2000,
  \nat, 404, 459

\bibitem[{{Raymond} \& {Smith}(1977)}]{Raymond77}
{Raymond}, J.~C., \& {Smith}, B.~W. 1977, \apjs, 35, 419

\bibitem[{{Sanders} {et~al.}(2001){Sanders}, {Edgar}, {Kraushaar}, {McCammon},
  \& {Morgenthaler}}]{Sanders2001}
{Sanders}, W.~T., {Edgar}, R.~J., {Kraushaar}, W.~L., {McCammon}, D., \&
  {Morgenthaler}, J.~P. 2001, \apj, 554, 694

\bibitem[{{Savage} \& {Sembach}(1996)}]{Savage1996}
{Savage}, B.~D., \& {Sembach}, K.~R. 1996, \araa, 34, 279

\bibitem[{{Smith} {et~al.}(2001){Smith}, {Brickhouse}, {Liedahl}, \&
  {Raymond}}]{Smith2001}
{Smith}, R.~K., {Brickhouse}, N.~S., {Liedahl}, D.~A., \& {Raymond}, J.~C.
  2001, \apjl, 556, L91

\bibitem[{{Smith} {et~al.}(2014){Smith}, {Foster}, {Edgar}, \&
  {Brickhouse}}]{Smith2014a}
{Smith}, R.~K., {Foster}, A.~R., {Edgar}, R.~J., \& {Brickhouse}, N.~S. 2014,
  \apj, 787, 77

\bibitem[{{Smith} {et~al.}(2007){Smith}, {Bautz}, {Edgar}, {Fujimoto},
  {Hamaguchi}, {Hughes}, {Ishida}, {Kelley}, {Kilbourne}, {Kuntz}, {McCammon},
  {Miller}, {Mitsuda}, {Mukai}, {Plucinsky}, {Porter}, {Snowden}, {Takei},
  {Terada}, {Tsuboi}, \& {Yamasaki}}]{Smith2007}
{Smith}, R.~K., {Bautz}, M.~W., {Edgar}, R.~J., {et~al.} 2007, \pasj, 59, 141

\bibitem[{{Snowden}(1993)}]{snowden93}
{Snowden}, S.~L. 1993, Advances in Space Research, 13,
  doi:10.1016/0273-1177(93)90102-H

\bibitem[{{Snowden} {et~al.}(1994){Snowden}, {McCammon}, {Burrows}, \&
  {Mendenhall}}]{Snowden1994a}
{Snowden}, S.~L., {McCammon}, D., {Burrows}, D.~N., \& {Mendenhall}, J.~A.
  1994, \apj, 424, 714

\bibitem[{{Snowden} {et~al.}(1995){Snowden}, {Freyberg}, {Plucinsky},
  {Schmitt}, {Truemper}, {Voges}, {Edgar}, {McCammon}, \&
  {Sanders}}]{Snowden1995a}
{Snowden}, S.~L., {Freyberg}, M.~J., {Plucinsky}, P.~P., {et~al.} 1995, \apj,
  454, 643

\bibitem[{{Snowden} {et~al.}(1997){Snowden}, {Egger}, {Freyberg}, {McCammon},
  {Plucinsky}, {Sanders}, {Schmitt}, {Tr{\"u}mper}, \& {Voges}}]{Snowden1997}
{Snowden}, S.~L., {Egger}, R., {Freyberg}, M.~J., {et~al.} 1997, \apj, 485, 125

\bibitem[{{Snowden} {et~al.}(2014){Snowden}, {Chiao}, {Collier}, {Porter},
  {Thomas}, {Cravens}, {Robertson}, {Galeazzi}, {Uprety}, {Ursino},
  {Koutroumpa}, {Kuntz}, {Lallement}, {Puspitarini}, {Lepri}, {McCammon},
  {Morgan}, \& {Walsh}}]{Snowden2014}
{Snowden}, S.~L., {Chiao}, M., {Collier}, M.~R., {et~al.} 2014, \apjl, 791, L14

\bibitem[{{Thomas} {et~al.}(2013){Thomas}, {Carter}, {Chiao}, {Chornay},
  {Collado-Vega}, {Collier}, {Cravens}, {Galeazzi}, {Koutroumpa}, {Kujawski},
  {Kuntz}, {Kuznetsova}, {Lepri}, {McCammon}, {Morgan}, {Porter}, {Prasai},
  {Read}, {Robertson}, {Sembay}, {Sibeck}, {Snowden}, {Uprety}, \&
  {Walsh}}]{Thomas2013}
{Thomas}, N.~E., {Carter}, J.~A., {Chiao}, M.~P., {et~al.} 2013, in \procspie,
  Vol. 8859, UV, X-Ray, and Gamma-Ray Space Instrumentation for Astronomy
  XVIII, 88590Z

\bibitem[{{Uprety}(2015)}]{Uprety2015}
{Uprety}, Y. 2015, PhD thesis, University of Miami

\bibitem[{{Welsh} \& {Shelton}(2009)}]{WelshShelton2009}
{Welsh}, B.~Y., \& {Shelton}, R.~L. 2009, \apss, 323, 1

\bibitem[{{Yoshino} {et~al.}(2009){Yoshino}, {Mitsuda}, {Yamasaki}, {Takei},
  {Hagihara}, {Masui}, {Bauer}, {McCammon}, {Fujimoto}, {Wang}, \&
  {Yao}}]{Yoshino2009}
{Yoshino}, T., {Mitsuda}, K., {Yamasaki}, N.~Y., {et~al.} 2009, \pasj, 61, 805

\end{thebibliography}

\end{document}